\documentclass{aastex}
\usepackage{emulateapj5}

\newcommand{\bvec}[1]{\mbox{\boldmath$#1$}}

\begin{document}
\title{Magnetorotational Instability\\ in Liquid Metal Couette Flow}
\author{K. Noguchi}
\affil{T-CNLS, Los Alamos National Laboratory, Los Alamos, NM 87545}
\author{V. I. Pariev\altaffilmark{1}\altaffilmark{2} and S. A. Colgate}
\affil{T-6, Los Alamos National Laboratory, Los Alamos, NM 87545}
\altaffiltext{1}{Lebedev Physical Institute, Leninsky Prospect 53,
Moscow 117924, Russia}
\altaffiltext{2}{Currently at Department of Physics and Astronomy,
University of Rochester, Rochester, NY 14627}
\author{J. Nordhaus}
\affil{Department of Physics and Astronomy, University of Rochester,
      Rochester, NY 14627}
\and
\author{H.F. Beckley}
\affil{Department of Physics, New Mexico Institute of Mining and
Technology, Socorro, NM 87801}

\begin{abstract}

Despite the importance of the
magnetorotational instability (MRI) as a  fundamental mechanism for
angular momentum transport   in magnetized accretion disks, it has yet
to be  demonstrated in the laboratory.  A liquid sodium
$\alpha\omega$ dynamo experiment at the New Mexico Institute of Mining
and Technology provides an ideal environment to study
the MRI in a rotating metal annulus (Couette flow).  A local stability
analysis is performed as a function of shear, magnetic field strength,
magnetic Reynolds number, and turbulent Prandtl number. The later
takes into account the minimum  turbulence induced by  the formation
of an Ekman layer against the rigidly rotating end walls of a cylindrical
vessel. Stability conditions are presented and unstable conditions for
the sodium experiment are compared with another proposed MRI
experiment with liquid gallium. 
Due to the relatively large magnetic
Reynolds number achievable in the sodium experiment, it should be
possible to observe the excitation of the MRI for a wide range of
wavenumbers and further  to observe the transition to the turbulent
state.
\end{abstract}
\nopagebreak

\section{Introduction}\label{sec1}

A significant problem in accretion disk theory is the nature of
anomalous viscosity. In order for accretion to occur, angular momentum
must be  transported outward.   The central problem in astrophysical
accretion disks is that observed  accretion rates cannot be due to
ordinary molecular viscosity. A robust anomalous angular momentum
transport mechanism must operate in accretion disks.

In 1991, the magnetorotational instability (MRI), discovered by
\citet{vel59} and \citet{cha60}, was reintroduced as a mechanism for
excitation and sustaining MHD turbulence  in a magnetized but
Rayleigh-stable fluid by \citet{bal91a}.   Since then, many numerical
and analytic studies of the MRI have been performed under varying
conditions
\citep{bal91b,mat95,haw96,sto96,gam96,san99,nog00, san01}.
Nevertheless, amidst all the theoretical attention granted to  the
MRI, it has never been demonstrated in the laboratory.  In light of
this fact an $\alpha \omega$ dynamo experiment at the New Mexico
Institute of Mining and Technology provides a unique opportunity to
study the MRI in a rotating metal annulus using liquid sodium. In this
paper, a local stability analysis is  performed and the results are
compared with theoretical analysis from a similar proposed experiment
at the Princeton Plasma Physics Laboratory \citep{ji01, go01}. Varying
aspects of the experiments are discussed with stable  and unstable
regions identified in terms of magnetic field strength and shear
flow.  In addition the number of unstable modes and the Prandtl number
further define the parameter space.  If the number of unstable modes
is large compared to unity, then there exists the possibility of
observing turbulence generated by the MRI. Finally we investigate the
instability boundary when fluid turbulence is injected as for example
through the Ekman layer flow.

\section{New Mexico $\alpha\omega$ Dynamo Experiment}
\label{sec2}

The New Mexico $\alpha\omega$ dynamo (NMD) experiment is a
collaboration between the New Mexico Institute of Mining and
Technology and Los Alamos National Laboratory \citep{col01b}. The
experiment is designed to create an astrophysical dynamo, the $\alpha
\omega$-dynamo, in a rapidly rotating laboratory system. 
\begin{figure*}[htb]
\plotone{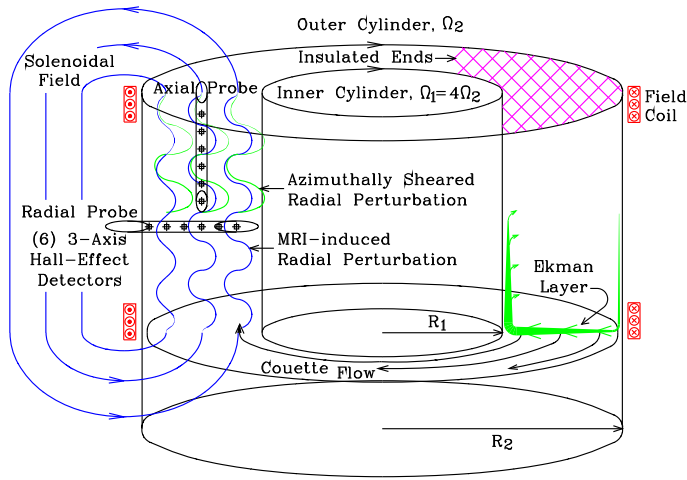}
\caption {A schematic drawing of  the New Mexico Institute of Mining and
Technology, $\Omega$-Phase, of the $\alpha \omega$ dynamo experiment.  The
inner cylinder is rotated relative to the outer or main cylinder of radius,
$R_{2} = 30.5$ cm at $\Omega_2$ and  where the inner cylinder of $R_1 =
15.2$ cm is rotated at $\Omega_1$ .  The ratio, $\Omega_1 /
\Omega_2$,  can be adjusted from unity to infinity by a brake on
$\Omega_2$, not shown, and  where $\Omega_1$ is driven.  A fixed gear ratio
also allows the maximum shear case at marginal stability,
$\Omega_1 / \Omega_2 = (R_2 / R_1)^2 =4$ to be obtained.  An axial or
quadrupole magnetic field is supplied by coils  shown also schematically.
The magnetic field internal to the liquid sodium in the annular space
between the two cylinders and bounded by the end walls (co-rotating with
the outer cylinder)  is measured by an aerodynamically shaped probe
containing six, 3-axis, Hall detectors.  These detectors can measure
fields from $0.1\,\mbox{G}$, the earth's field, to up 
$10\,\mbox{kG}$. Thus the initial static
magnetic  field distribution,  $B_z$, can be  measured  and
compared to the  field components, $B_r$ and $B_{\theta}$, produced by the
MRI.  The probe can be mounted either radially or axially depending upon
the mode information desired.}
\label{fig1}
\end{figure*}
The apparatus
consists of two coaxial cylinders (Fig.~\ref{fig1}), rotating at
different angular velocities and therefore creating Couette flow in
the annular volume.  Liquid sodium  fills the volume  between the
cylinders and  the end walls. Solid plates attached to and co-rotating
with the outer cylinder with an angular velocity, $\Omega_2$ define
the end walls. (In addition, for the dynamo experiment, an external
source of helicity  is supplied, driven plumes,  but this is not part of
the MRI experiment.) The  schematic of the flow field, 
(Fig.~\ref{fig1}), places particular emphasis on the primary diagnostic of
multiple, 3-axis, magnetic field Hall effect detectors (sensitivity: $0.1$
to $10\,\mbox{kG}$) located in aerodynamically shaped probes 
within the rotating conducting fluid.  
We expect that the radial perturbations from the
MRI and their azimuthally sheared result  will produce a   fluctuating
$B_r$ and $B_{\theta}$ field from an original imposed static $B_z$
field through MRI growth.  
These fluctuating fields are the result of
the linear and non-linear growth of 
the various MRI modes transformed by the difference of the sheared Couette flow
at a given radius and  the probe angular  velocity, $\Omega_2$, of the
outer cylinder.  
A significant difficulty will be the observation of the
linear growth  of any particular MRI mode because the time constant for
establishing the initial axial field within the conducting liquid sodium
will be long,
$\sim 30/\Omega_2$, compared to the expected growth rate, $\sim
\Omega_2$, of the instabilities as derived in this paper.  We therefore
expect to observe primarily the near steady state of the non-linear
limit of  various modes, but the sequential linear phases may be observed
during the comparatively slow rise of the field.  If the applied field
or flux is amplified by the MRI such as a dynamo, then we expect to see
fluctuating fields  significantly greater than the applied field.  In
addition since the inner and outer cylinders are driven separately, the
relative torque as a function of the applied magnetic field becomes an
integral diagnostic of the non-linear limits of the instability growth.

  By driving
the inner cylinder and applying a variable brake with a corresponding
torque measurement to the outer cylinder  one can explore the
full range of Couette velocity profiles  including the marginal Couette
flow hydrodynamic stability condition discussed next. 
This condition of maximum or
marginal  stable Couette profile  can be established in the  experiment
precisely by gear ratios and so the degree of turbulence measured by the
torque can be explored at the stability boundary.  In addition the
pressure will be measured at five radii  and compared to the pressure
distributions expected  of the various Couette profiles.  A finite torque
measurement can be interpreted in terms of turbulence existing between the
two cylinders.  No turbulence or perfectly laminar flow will  exert a
torque of the order
$1/R_e$, $R_e$ the fluid Reynolds number where  $Re \simeq 10^7$,  compared to
a turbulent torque, $\sim 1/Re^{1/2}$,   if the Ekman layer
circulation leads to the weak turbulence that we discuss  later. This same
possible weak turbulence can also be measured by introducing a very weak
field, $B_{min} \simeq 1\,\mbox{G}$, small enough so as not to 
cause the growth of MRI in resistive liquid
but large enough so that an unstable flow or weakly
turbulent flow can be measured as fluctuations in $B_r$ and $B_{\theta}$ with
the Hall effect probes.  Therefore the fluid flow conditions 
can be fully explored before the application of magnetic fields 
designed to create the MRI. When
the MRI does take place, then the instability can be recognized as a
departure from the previously measured initial fluid state.

It is critical to have large shear rates in order to observe the maximum
growth rates of the MRI.
However, excessive shear will hydrodynamically
destabilize the flow by the Kelvin-Helmholtz instability.  Let us
consider  a Couette flow profile in cylindrical coordinates.  Take
$r, \theta, z$  as the radial, azimuthal and axial directions
respectively.  The radial distribution of angular velocity of the flow,
$\Omega ( r )$, is given by \citep{lan59}
\begin{equation}
\Omega(r)=\frac{\Omega_2{R}_2^2-\Omega_1{R}_1^2}{{R}_2^2-{R}_1^2}+
\frac{1}{r^2}\frac{(\Omega_1-\Omega_2){R_1^2}{R_2^2}}{{R}_2^2-{R}_1^2}
\mbox{,}
\end{equation}  where $R_1(R_2)$ and $\Omega_1(
\Omega_2)$ are the inner(outer) radii and angular velocities.

In the limit of infinitely large hydrodynamic Reynolds number,
$R_{e}$, the stability condition  for Couette flow is given by
$\Omega_1R_1^2$ $<$ $\Omega_2R_2^2$  \citep{lan59}.
      Therefore, in order to maximize the shear flow within the
apparatus,  the NMD experiment has been designed such that
${R_2}/{R_1} = 2$ and
$\Omega_1/\Omega_2$ = 4, guaranteeing that
$\Omega_1R_1^2$ = $\Omega_2R_2^2$.
In addition to stability constraints, stress
limitations in the experiment require that an upper limit of
${\Omega_2} = 33\,\mbox{Hz}$ be placed on the frequency of rotation of the
outer cylinder. Of course lower rotation rate can be used in 
both NMD experiment and Princeton Plasma Physics Laboratory (PPPL) 
experiment,
but it is assumed for this analysis that the
highest rates are of greatest scientific interest.

The assumption of stable Couette flow implies a laminar flow with no
turbulence. On the other hand the initial acceleration of the fluid to
the final state of Couette flow from an alternate initial state
implies a transient enhanced torque, because, just as in the accretion
disk, the laminar friction is too small.  However, in the experimental
apparatus, the transient, Couette flow profile is Helmholtz
unstable so that turbulence is a natural  and expected result of the
"spin-up" of the flow.

On the other hand the  Ekman flow creates a relative torque between
$\Omega_1$ and $\Omega_2$ that we expect to be balanced by a weak
turbulence as observed by \citep{tay36} and analogous to the spin-up
turbulence.   This  turbulence may also influence  the
stability conditions but primarily the ability to distinguish
turbulence  caused by the MRI from the hydrodynamic turbulence caused
by the Ekman layer.  We therefore analyze the MRI stability conditions
as a function of hydrodynamic turbulence
preexisting in the  liquid  and therefore of the Prandtl number. At
large enough levels of turbulence, the effective electrical
resistivity can also be increased and therefore decrease the magnetic
Reynolds number, $R_m$, and therefore influence the conditions of
excitation of the MRI.

In comparison to the New Mexico Dynamo Experiment, a similar
experiment  at the Princeton Plasma Physics Laboratory has been
proposed to look for the  MRI in a rotating liquid metal annulus
\citep{ji01}.   The  PPPL experiment utilizes liquid gallium, an easy
to handle metal with properties similar to liquid sodium 
(see Table~1).
\begin{table*}
\begin{center}
\caption{Actual and Normalized Quantities in the Sodium and  Gallium
Experiments}
\begin{tabular}{lcccc}
\tableline\tableline
&\multicolumn{2}{c}{Actual}&\multicolumn{2}{c}{Normalized}\\ Property&
Sodium & Gallium & Sodium & Gallium \\
      \tableline Kinematic Viscosity, $\nu$(${cm^2} {s^{-1}}$) &
7.1$\cdot10^{-3}$ &  3.2$\cdot{10^{-3}}$& 3.6$\cdot{10^{-8}}$ &
2.2$\cdot{10^{-7}}$ \\ Reynolds Number, $R_e$
&-&-&$1.3\cdot10^7$&$3.0\cdot10^6$\\ Magnetic Diffusivity,
$\eta$(${cm^2} {s^{-1}}$) & 810 & 2000 & 4.2$\cdot{10^{-3}}$ &
1.4$\cdot{10^{-1}}$\\ Magnetic Reynolds Number,
$R_m$&-&-&120&4.7\\ Density, $\rho$(g cm$^{-3}$) & 0.92 & 6.0& - & -
\\ Alfv\'en Speed, ${V_A}$(cm s$^{-1}$) ($10^3$\,\mbox{gauss}) &
$2.9\cdot10^2$&
$1.1\cdot10^2$& 4.6$\cdot{10^{-2}}$ & 1.2$\cdot{10^{-1}}$
\\ Inner Radius, ${R_1}$(cm) & 15.25 & 5 & $.5$ &
$.33$\\ Outer Radius, ${R_2}$(cm) & 30.5 & 15 & 1 & 1\\ Length,
$L$(cm) & 30.5&10.&1&0.66\\ Inner Angular Velocity,
$\Omega_1$(s$^{-1})$ & 829& 533& 4 & 8.2\\ Outer Angular Velocity,
$\Omega_2$(s$^{-1})$ & 207& 65& 1 & 1\\ Prandtl Number,
$P_M=R_m/R_e$ &-&-&$9.2\cdot10^{-6}$&$1.6\cdot10^{-6}$\\ Ekman
Turbulent Prandtl \#,
$P_{Mt}$&-&-&$0.012$&$6.3\times10^{-3}$\\
\tableline
\end{tabular}
\end{center}
\end{table*}
Note, however, the higher  density
and higher resistivity of liquid gallium, which limit the maximum
rotation speed and the maximum achievable
$R_m$. The dimensions of the PPPL experiment are slightly different,
enabling  them to acquire larger shear flow rates,  \citet{ji01}.
For ${R_1} = 5$ cm and
${R_2} = 15$ cm then
${{R_2}/{R_1}}$ = 3 with a typical
$\Omega_1/\Omega_2$ = 9. 
The conditions for instability  for both
experiments are discussed in Sec.~4.

\section{Local Stability Analysis}\label{sec3}

The angular velocity of Couette flow confined between coaxial
cylinders with radii $R_1 < r<R_2$ and cylindrical angular velocities
$\Omega_1,\Omega_2$ is given by
\begin{equation}
\Omega(r)=a+\frac{b}{r^2}\label{Couette},
\end{equation}  where we define $a$ and $b$ as
\begin{eqnarray}
a&=&\frac{\Omega_2{R}_2^2-\Omega_1{R}_1^2}{{R}_2^2-{R}_1^2},\nonumber\\
b&=&\frac{(\Omega_1-\Omega_2){R_1^2}{R_2^2}}{{R}_2^2-{R}_1^2}.\label{ab}
\end{eqnarray}

The incompressible and dissipative MHD equations describing the
dynamics of liquid metals are given as follows,
\begin{mathletters}
\begin{eqnarray}
\nabla\cdot\bvec{B}&=&0,\label{b1}\\
\nabla\cdot\bvec{V}&=&0,\\
\frac{\partial \bvec{B}}{\partial
t}&=&\nabla\times(\bvec{V}\times\bvec{B})+\eta\nabla^2\bvec{B},\\
\frac{\partial\bvec{V}}{\partial
t}+(\bvec{V}\cdot\nabla)\bvec{V}&=&\frac{(\bvec{B}\cdot\nabla)\bvec{B}}{4\pi\rho}-\frac{1}{\rho}\nabla\left(p+\frac{B^2}{8\pi}\right)\nonumber\\
&&\quad+\nu\nabla^2\bvec{V}\label{b4},
\end{eqnarray}
\end{mathletters} where $\bvec{B}$ is the magnetic field,
$\bvec{V}$ is the velocity,
$\eta$ is the magnetic diffusivity, $p$ is pressure and
$\nu$ is the kinematic viscosity.
In cylindrical symmetry the system of equations
(\ref{b1})--(\ref{b4}) have stationary solution
$\bvec{V}_0=(0,r\Omega(r),0)$ and
$\bvec{B}_0=(0,B_{\theta 0}(r),B_{z 0})$, where $B_{z0}$ is a constant,
$B_{\theta 0}\propto 1/r$, and the angular velocity profile,
$\Omega(r)$, is given by expression~(\ref{Couette}).
One needs to investigate the
time evolution of perturbations to this  equilibrium state governed by
the linearization of system of  equations (\ref{b1})--(\ref{b4}).
A similar analysis of perturbations was performed by \citet{go01}, who
showed that local WKB approximation gives results for growth rates of
instability, which are close to the growth rates obtained by the
solving full boundary value problem in radial  direction. As it is
especially stressed by
\citet{ji01} and \citet{go01}, WKB local analysis leads to a good
approximation to the growth rates even in the case of the scale of
perturbations being equal to or  comparable to the sizes of the
vessel. Thus, in this work, we limit  ourselves to the local approach,
which is much easier to carry out than the full eigenmode analysis,
because it allows one to obtain an algebraic  dispersion relation.
The perturbations
$\bvec{b}=(b_r,b_\theta,b_z)$,
$\bvec{v}=(v_r,v_\theta,v_z)$ are assumed to be axisymmetric and
proportional to
$\exp(\gamma{t}-i{k_z}z-i{k_r}r)$ where
$\gamma$ is the associated growth rate.  It is also assumed that the
minimum possible wave numbers in $r$ and $z$ directions are
$k_{rmin}={\pi}/{(R_2-R_1)}$ and
$k_{zmin}={\pi}/{L}$. The linearized equations of motion are then
given by
\begin{mathletters}
\begin{eqnarray} 0&=&\left(\frac1{r}-ik_r\right)v_r-ik_zv_z,\\
0&=&\left(\frac1{r}-ik_r\right)b_r-ik_zb_z,\\
\gamma b_r&=&-ik_zB_{z0}v_r-\eta k^2b_r,\\
\gamma b_\theta&=&-ik_zB_{z0}v_\theta+\frac{d\Omega}{d\ln
r}b_r\nonumber\\
&&\quad-rv_r\frac{d}{dr}\left(\frac{B_{\theta0}}{r}\right)-\eta
k^2b_\theta,\\
\gamma v_r-2\Omega v_\theta&=&-\frac{1}{4\pi\rho}\left(ik_zB_{z0}b_r
+\frac{2B_{\theta0}}{r}b_\theta\right)\nonumber\\
&&\quad+ik_r\frac{p_1}{\rho}-\nu
k^2v_r,\\
\gamma v_\theta+\frac{\kappa^2}{2\Omega}v_r&=&-\frac{1}{4\pi\rho}
\left(ik_zB_{z0}b_\theta-b_r \left[\frac{d}{d
r}+\frac1{r}\right]B_{\theta0}\right)\nonumber\\
&&\quad-\nu k^2v_\theta,\\
\gamma v_z&=&-\frac{ik_zB_{z0}}{4\pi\rho}b_z+ik_z\frac{p_1}{\rho}-\nu
k^2v_z,
\end{eqnarray}
\end{mathletters} where the epicyclic frequency $\kappa$ is defined as
\begin{equation}
\kappa^2=\frac{1}{r^3}\frac{d({r^4}\Omega^2)}{dr}=4\Omega^2+\frac{d\Omega^2}{d\ln
r},
\end{equation}
$p_1$ is the perturbation of the pressure, and
$k=\sqrt{k_z^2+k_r^2}$ is total wave number, respectively. Note that
$\kappa$ can be expressed through $a$ and $b$ in Eq.~(\ref{ab}) as
\begin{equation}
\kappa^2=4a\left(a+\frac{b}{r^2}\right),\label{c}
\end{equation} which vanishes when velocity shear is maximum ($a=0$,
$b=r^2\Omega$).

These equations lead to the following local dispersion relation
\begin{eqnarray}
\lefteqn{\left[(\gamma+\nu{k^2})(\gamma+\eta{k^2})+{{k_z}^2{V_A}}^2\right]^2\frac{k^2}{k_z^2}}\nonumber\\
&&+\kappa^2(\gamma+\eta{k^2})^2+\frac{d\Omega^2}{d{\ln
r}}{{k_z}^2{V_A}}^2\nonumber\\
&&+\frac{2i{k_z}{V_{A\theta{z}}^2}}{r}
\left[(\gamma+\nu{k^2})\frac{d\Omega}{d{\ln
r}}-\frac{\kappa^2}{2\Omega}{(\gamma+\eta{k^2})}\right]=0,
\end{eqnarray} where
\begin{eqnarray}
{V_{A\theta{z}}}^2&=&\frac{B_{\theta{0}}B_{z0}}{4\pi\rho},
\quad V_A^2=\frac{B_{z0}^2}{4\pi\rho}
\end{eqnarray}

Neglecting  all $1/r$ terms compared to $k$ yields the following
dispersion relation, which is identical to the dispersion relation
derived by \citet{ji01}
\begin{eqnarray}
\lefteqn{[(\gamma+\nu{k^2})(\gamma+\eta{k^2})+({k_z}{V_A})^2]^2\frac{k^2}{k_z^2}}\nonumber\\
&&+\kappa^2(\gamma+\eta{k^2})^2+\frac{d\Omega^2}{d{\ln
        r}}({k_z}{V_A})^2=0.\label{dis}
\end{eqnarray}

In the case of maximum shear flow, $a=0$, and hence,
$\kappa=0$,
the dispersion relation simplifies to
\begin{equation} [(\gamma+\nu{k^2})(\gamma+\eta{k^2})+({k_z}{V_A})^2]^2
-4\Omega^2\frac{k_z^4V_A^2}{k^2}=0,
\label{dis2}
\end{equation} which immediately yields the following solutions for
$\gamma$
\begin{eqnarray}
\lefteqn{\gamma={\frac{1}{2}}\bigg[{-(\nu+\eta){k^2}}}\nonumber\\
&&\pm\sqrt{(\nu+\eta)^2
k^4-4\left(\nu\eta{k^4}+{k_z^2}{V_A^2}\pm\frac{2\Omega{k_z^2}{V_A}}{k}
          \right)}\bigg],\label{sol}
\end{eqnarray}  
Only when we take the plus sign for the square root
term and the minus sign for the last term in eq. (\ref{sol}), does it
give the unstable solution, and all the other three solutions are
stable.

The MRI occurs only when the second term inside the square  root of
(\ref{sol}) is negative, i.e.,
\begin{equation}
{\omega_\nu}{\omega_\eta}+{\omega_A^2}<{2\frac{{k_z}{\Omega}{\omega_A}}{k}},
\end{equation}  where ${\omega_\nu}=\nu{k^2}$,
${\omega_\eta}=\eta{k^2}$ and
${\omega_A}={V_A}{k_z}$.  Thus, viscosity, magnetic diffusion and
magnetic tension stabilize the MRI, whereas the shear flow
destabilizes it.  The condition for neglecting
$\nu$ and $\eta$  can be derived from (\ref{sol}) by evaluating the
expression under the square root, i.e.,
\begin{equation}
(\nu-\eta)^2{k^4}{\ll}\frac{4\omega_A(2{k_z}{\Omega}-{k}{\omega_A})}{k}.
\label{sol2}
\end{equation} For ${k_r}=0$, eq. (\ref{sol2}) further reduces to
\begin{equation}
\frac{(\eta-\nu)^2{k_z^2}}{V_A^2}\ll\frac{4(2\Omega-\omega_A)}{\omega_A}.
\end{equation}  Thus, it is apparent that magnetic diffusivity and
kinematic viscosity only affect high
${k_z}$ modes.

If the magnetic diffusivity is large and the applied magnetic field is
weak,  eq. (\ref{sol}) reduces to two roots
${\gamma=-\nu{k^2}}$ and ${\gamma=-\eta{k^2}}$.   The former root
corresponds to a hydrodynamical branch in which the fluid  is
disconnected from the electromagnetic force and behaves as a pure
fluid. The latter root is an electromagnetic branch, in which the
magnetic field diffuses as in vacuum.  With an increasing magnetic
field, bifurcations occur in which the hydrodynamic and
electromagnetic branches are split into four branches.  When
$\omega_\nu\sim\omega_A\ll\omega_\eta$ the unstable solution of
(\ref{dis2}) is  given by
\begin{equation}
\gamma=-\omega_\nu+\frac{\omega_A{(2\Omega-\omega_A})}{\omega_\eta-\omega_\nu},
\end{equation} showing that the unstable solution emerges from the
hydrodynamical branch. Though magnetic diffusion diminishes the MRI,
the branch remains unstable if the condition
$\omega_A>\omega_\eta\omega_\nu k/(2k_z\Omega)$ is satisfied. Notice
however, that even if the magnetic diffusivity is high, a weak
magnetic field is capable of generating the MRI.

Next, let us consider the hydrodynamical limit($V_A=0$) with arbitrary
Couette flow profiles, ($a\neq0$). In this case,
$\kappa\neq0$ in general and we have to go back to eq. (\ref{dis}) for
deriving solutions. Two trivial solutions are
$\gamma=-\eta k^2$, corresponding to the electromagnetic branch and
the second is given by
\begin{equation}
\gamma=-\nu k^2\pm i\kappa\frac{k_z}{k},\label{sol4}
\end{equation} which is the hydrodynamical branch.

\begin{figure*}[htb]
\begin{tabular}{cc}
\hspace{5mm}\includegraphics[width=8cm]{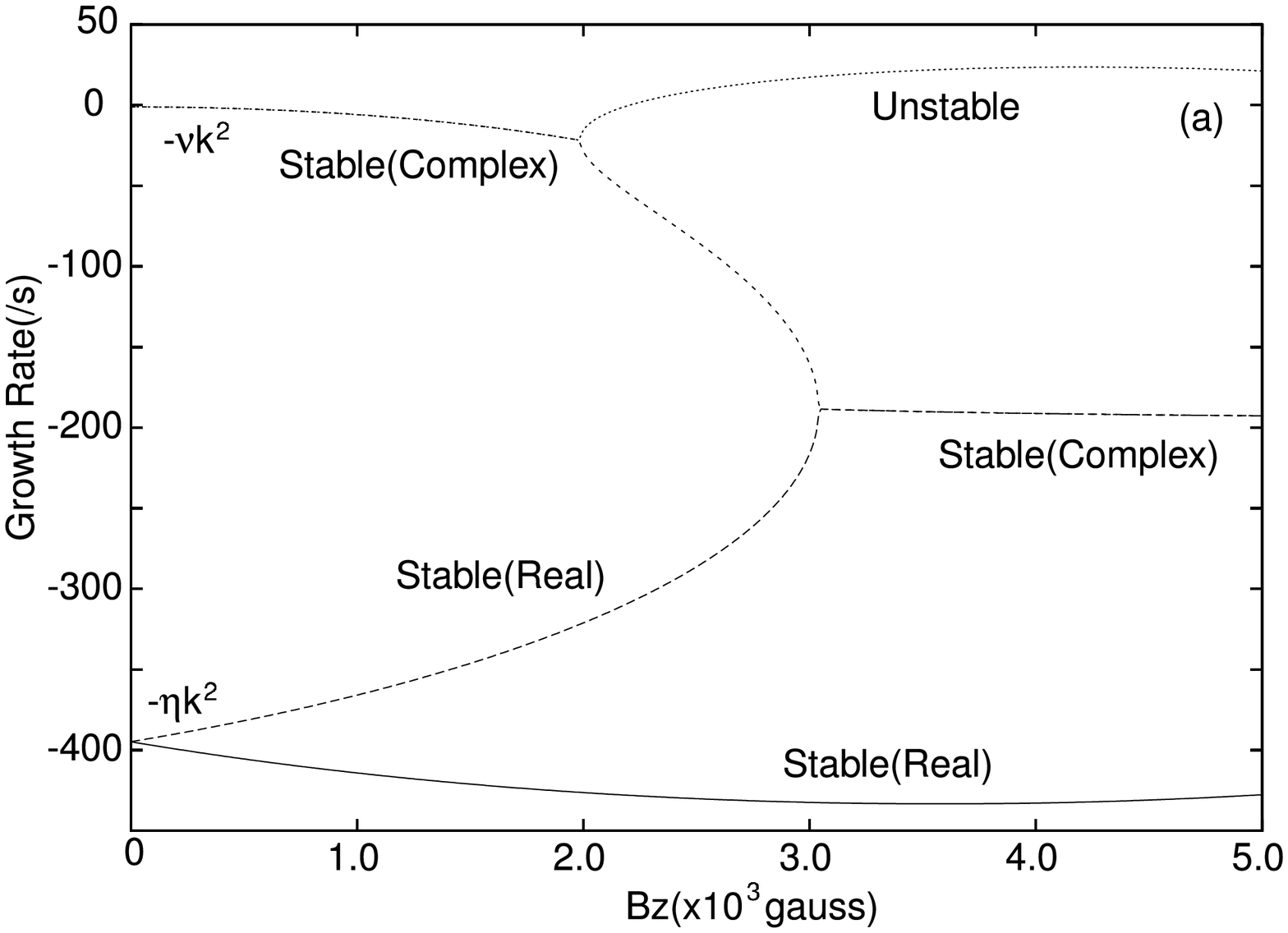}\hspace{5mm}
\includegraphics[width=8cm]{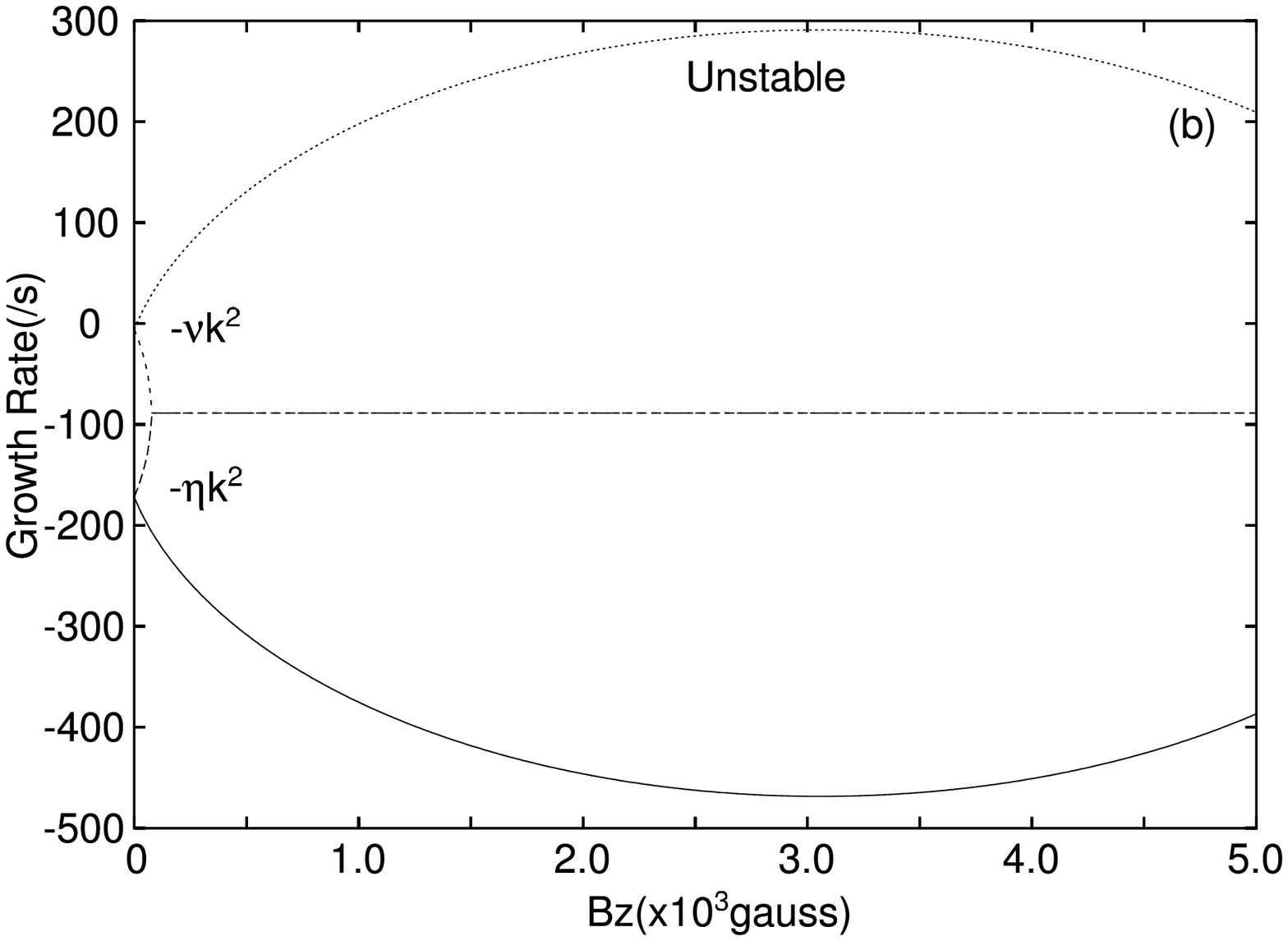}\\
\hspace{5mm}\includegraphics[width=8cm]{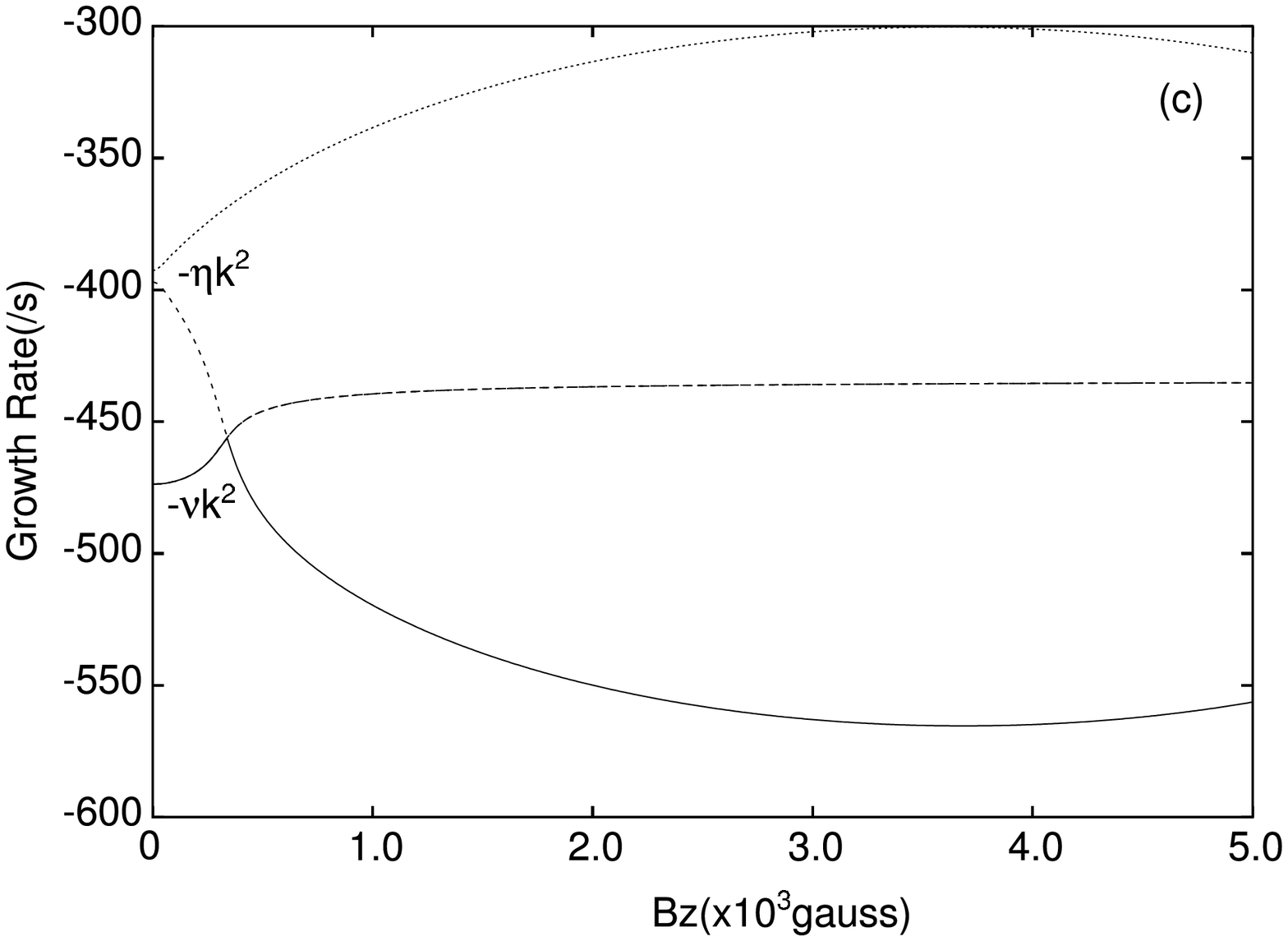}\hspace{5mm}
\includegraphics[width=8cm]{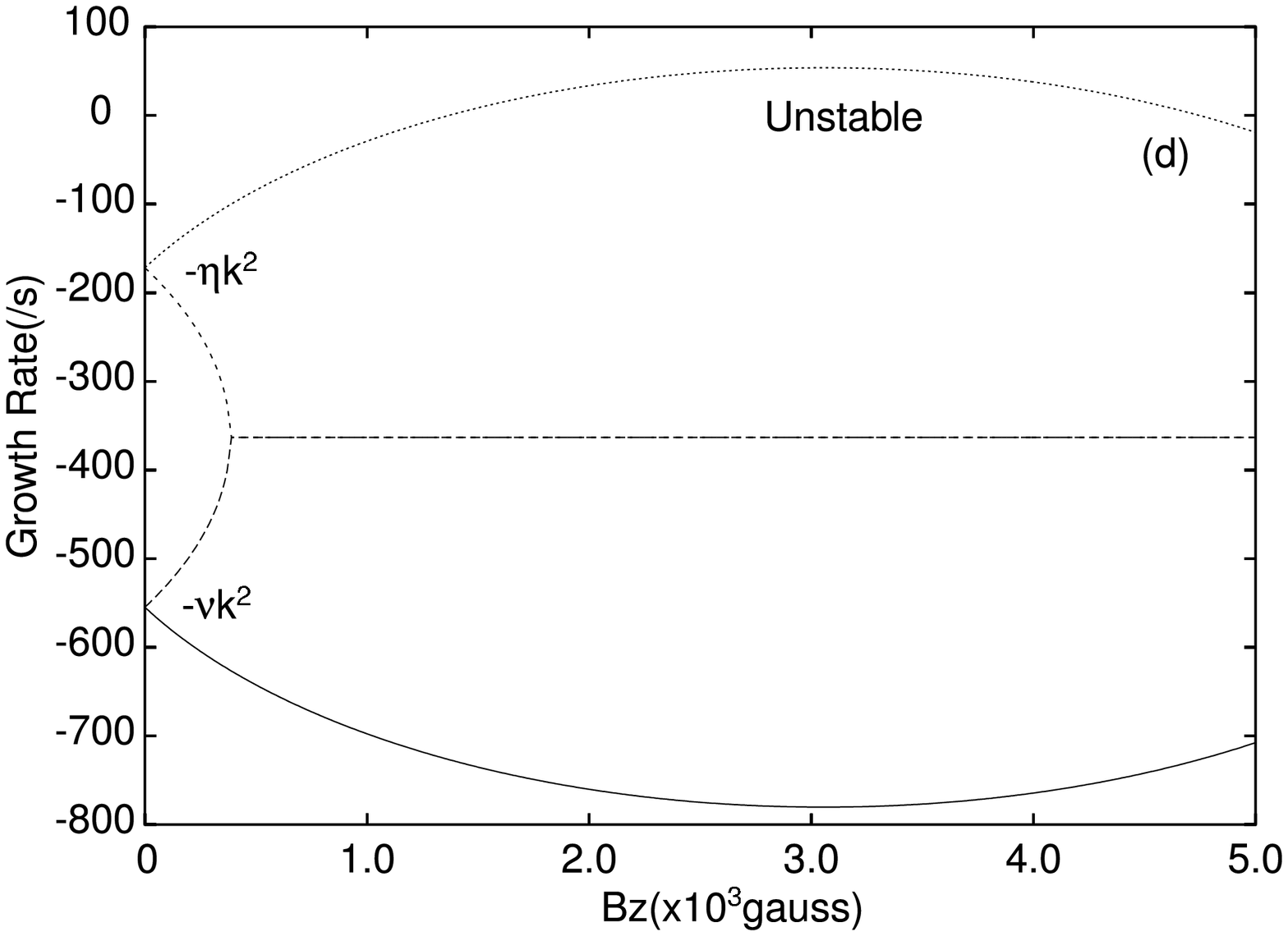}
\end{tabular}
\caption{The complex stability plane of the combinations of limiting
Couette and non-limiting Couette flow profiles  with laminar and
turbulent flows. Figs. (a) and (c) are for a non-limiting Couette flow
profile (the gallium experiment), and Figs. (b) and (d) are for the
limiting  Couette flow profile (the sodium experiments). In Figs. (a)
and (b), the flow is assumed as laminar($\nu\ll\eta$), and (c) and
(d), the flow is turbulent, $\nu>\eta$. The corresponding Prandtl
numbers are $P_{Mt}=1.20$ for (c), and
$3.23$ for (d). A remarkable difference between the non-limiting
Couette(a) and limiting Couette (b) profiles is the hydrodynamical
branch, where ($-\nu k^2$) is stable with the non-limiting Couette
profile and weak magnetic field, whereas it becomes unstable with a
weak magnetic field  and the limiting  Couette flow profile. With
strong turbulence, [Figs. (c) and (d)], the non-limiting Couette flow
profile is always stable (c), whereas the limiting Couette flow
profile becomes unstable with a strong magnetic field
($B_z>2\times10^3\,$gauss).
In Figs.~(a) and~(c)
$(k_z,k_r)=(1,1)$; in Figs.~(b)  and~(d)
$(k_z,k_r)=(4,1)$.}
\label{fig2}
\end{figure*}
The effect of finite
$\kappa$ and high $\nu$ is shown in Fig.~\ref{fig2},  where the
rotation speed of the cylinders, viscosity, and magnetic diffusivity of
Fig.~\ref{fig2}a corresponds to the point  C of \citet{ji01}, and the
wavenumber is fixed at $(k_z, r_r)=(1,1)$  for Fig.~\ref{fig2}a,
\ref{fig2}c and $(4,1)$ for Fig.~\ref{fig2}b,
\ref{fig2}d, respectively.
The growth rates of the four roots of eq.
(\ref{dis}) are shown as a function of the axial magnetic field
strength $B_z$. The epicyclic frequency
$\kappa$ is finite in Figs.~\ref{fig2}a and \ref{fig2}c, whereas
$\kappa=0$ in Figs.~\ref{fig2}b and \ref{fig2}d.  The kinematic
viscosity $\nu$ is taken as the actual value of Gallium
[Fig.~\ref{fig2}a] and Sodium [Fig~\ref{fig2}b], whereas we make it
artificially high ($\nu>\eta$) in Figs.~\ref{fig2}c and \ref{fig2}d to
see the effect of anomalous increase of $\nu$ due to possible
turbulence.

In the hydrodynamical limit($B_z=0$) in Fig.~\ref{fig2}a, the
solutions of the hydrodynamical branch, ($\gamma\sim0$) are complex
[eq.~(\ref{sol4})]. Near
$B_z=2000$ gauss, these solutions are separated and both become real.
Only one solution becomes unstable. Thus, if the flow is not a maximum
shear flow profile ($\kappa\neq0$), the MRI is stabilized for weak
magnetic fields.

Figures (\ref{fig2}c) and (\ref{fig2}d) show that the  turbulence
suppresses the unstable MRI mode. Since the Ekman layer may make the fluid
weakly turbulent, it is important to estimate the scale of this turbulence,
which will be discussed in Sec.~\ref{sec5}.

In the next section, stability diagrams are presented and compared for
both experiments.

\section{Stability Diagrams and Growth Rates in Sodium and Gallium
Experiments}\label{sec4}

The comparison between the NMD and Princeton experiments is done by
comparing their typical parameters in Table 1. In order to evaluate
the physical differences between the experiments, we compare the
dimensionless parameters, presented in the second column of Table 1.
We use $R_2$ and
$\Omega_2^{-1}$ as units of length and  time to obtain the
dimensionless quantities.

We choose $\Omega_1=84.8$ Hz and $\Omega_2= 10.34$ Hz as the typical
values for the  gallium experiment, which corresponds to point C of
\citet{ji01}. The global magnetic Reynolds number becomes
\begin{equation} R_m=\frac{R_2\Omega_2(R_2-R_1)}{\eta}.
\end{equation}    The $R_m$ is higher in the sodium experiment, which
is designed to observe the
$\alpha\omega$ dynamo
\citep{col01b, par01}. All the growth rates are evaluated at the radius
$r=\bar{r}$, which satisfies
$\Omega(\bar{r})=\sqrt{\Omega_1\Omega_2}$.  We also define the global
fluid Reynolds number as
\begin{equation} R_e=\frac{R_2\Omega_2(R_2-R_1)}{\nu}.
\end{equation}

\begin{figure*}[hbt]
\begin{tabular}{cc}
\hspace{5mm}\includegraphics[width=8cm]{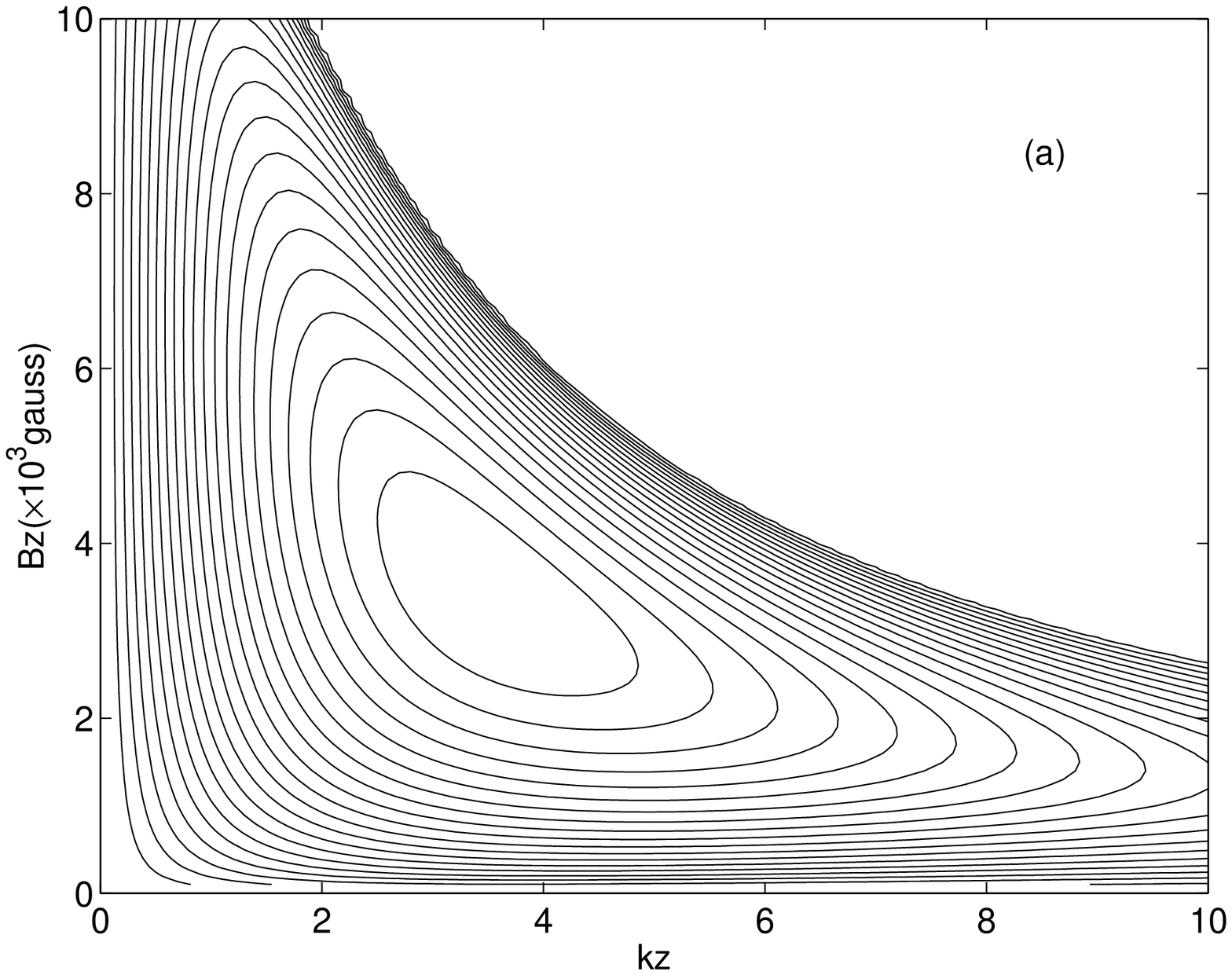}\hspace{5mm}
\includegraphics[width=8cm]{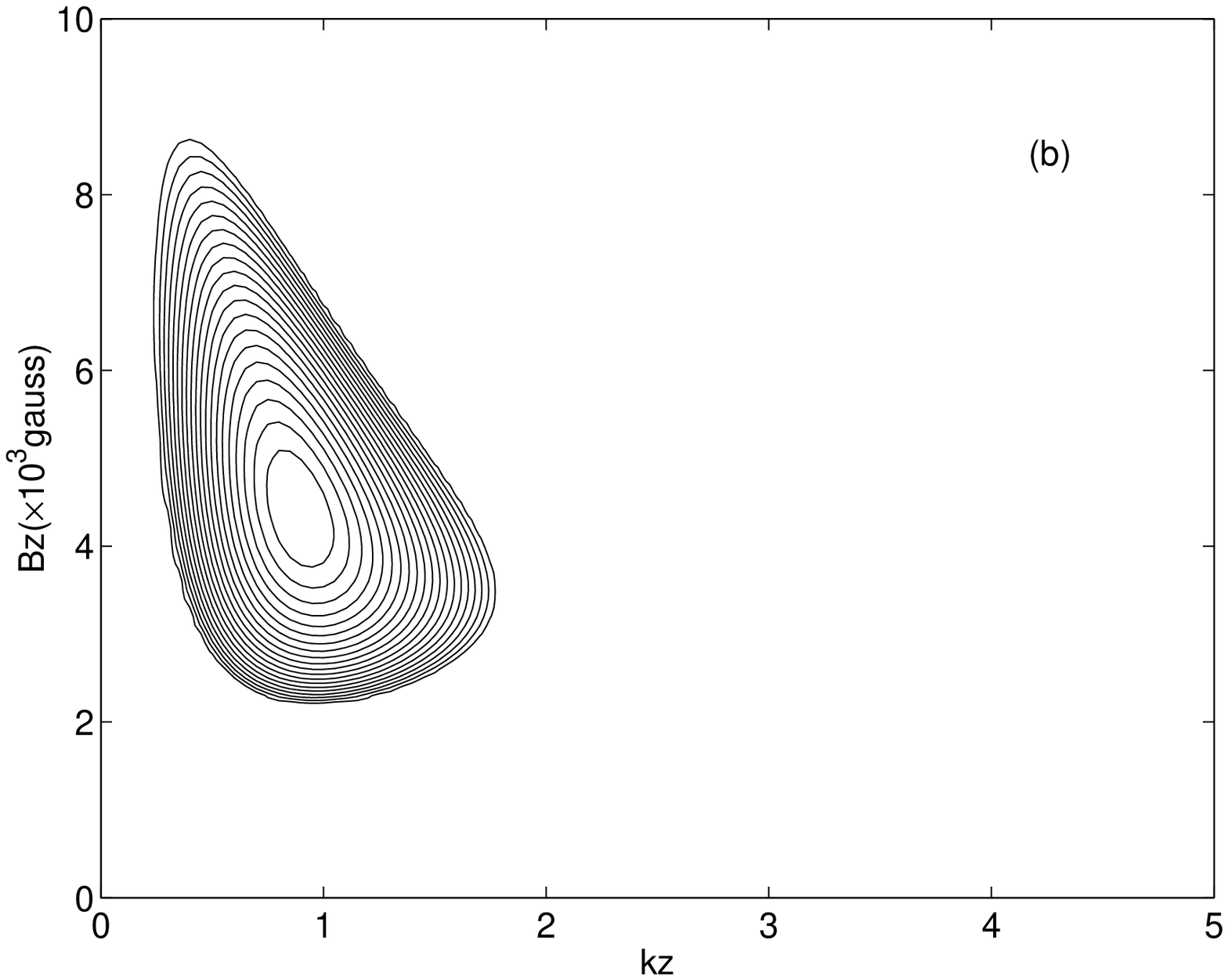}
\end{tabular}
\caption{Growth rates of the MRI in sodium (a) and gallium (b)
experiments
Growth rates are shown as a function of $k_z$ and
$B_z$. All the contours of contour figures are
equally spaced, and only positive growth rate contours are shown. The radial wavenumber $k_r$ is fixed to be unity for both
calculations. When $B_z=3\times10^3$gauss, $k_z=4$ mode is the most
unstable mode with $\gamma=300 s^{-1}$, and for the sodium
experiment(a), whereas $k_z=1$ with $\gamma=25 s^{-1}$  for the
gallium(b).  Higher
${k_z}$ modes are destabilized with weak magnetic field in the sodium
experiment,  but with strong magnetic field, they are suppressed by
the magnetic diffusivity and magnetic tension. Weak magnetic field
modes in gallium are suppressed because of finite $\kappa$ 
(see Fig.~\ref{fig51})}
\label{fig3}
\end{figure*}
\begin{figure*}[hbt]
\begin{tabular}{cc}
\hspace{5mm}\includegraphics[width=8cm]{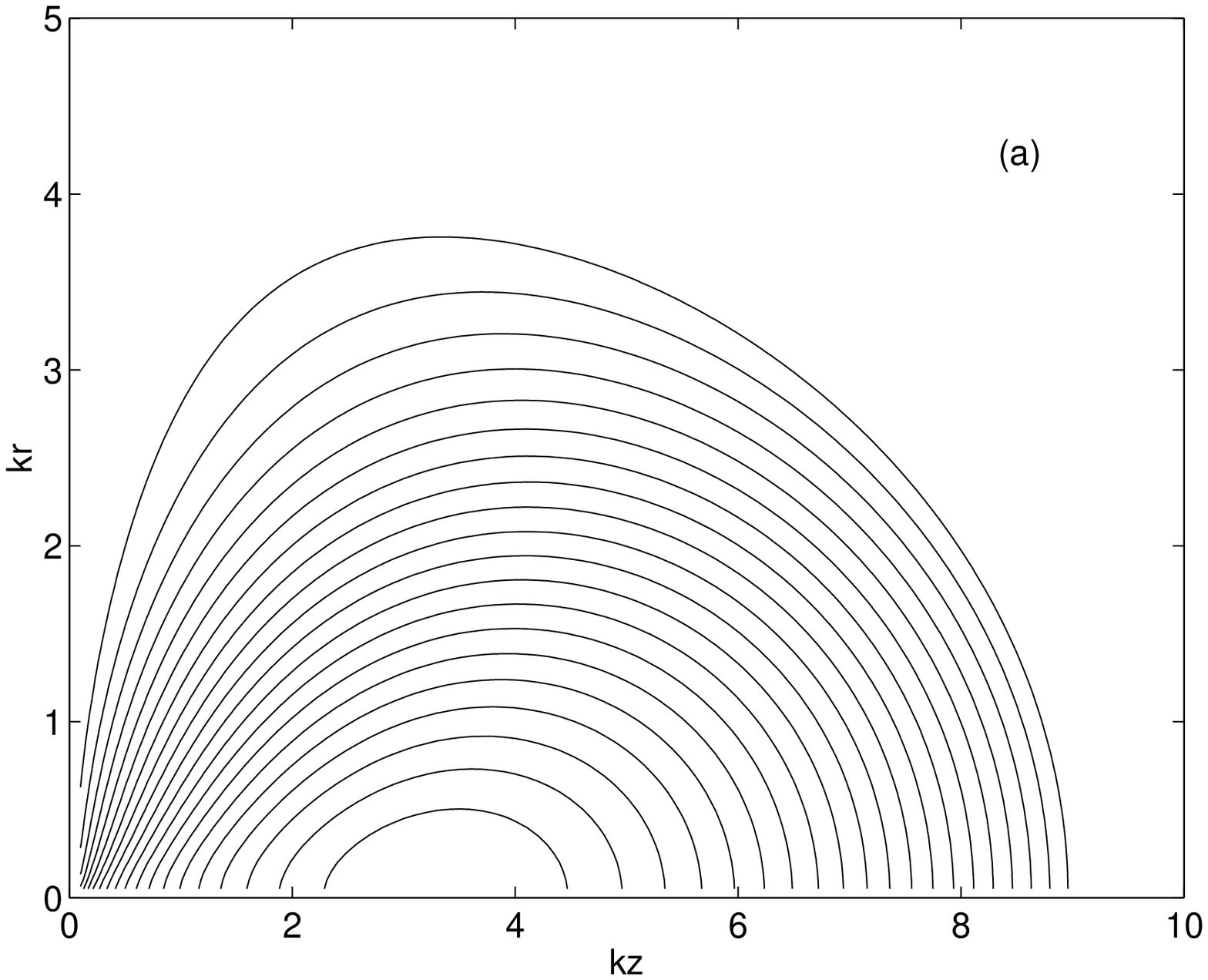}\hspace{5mm}
\includegraphics[width=8cm]{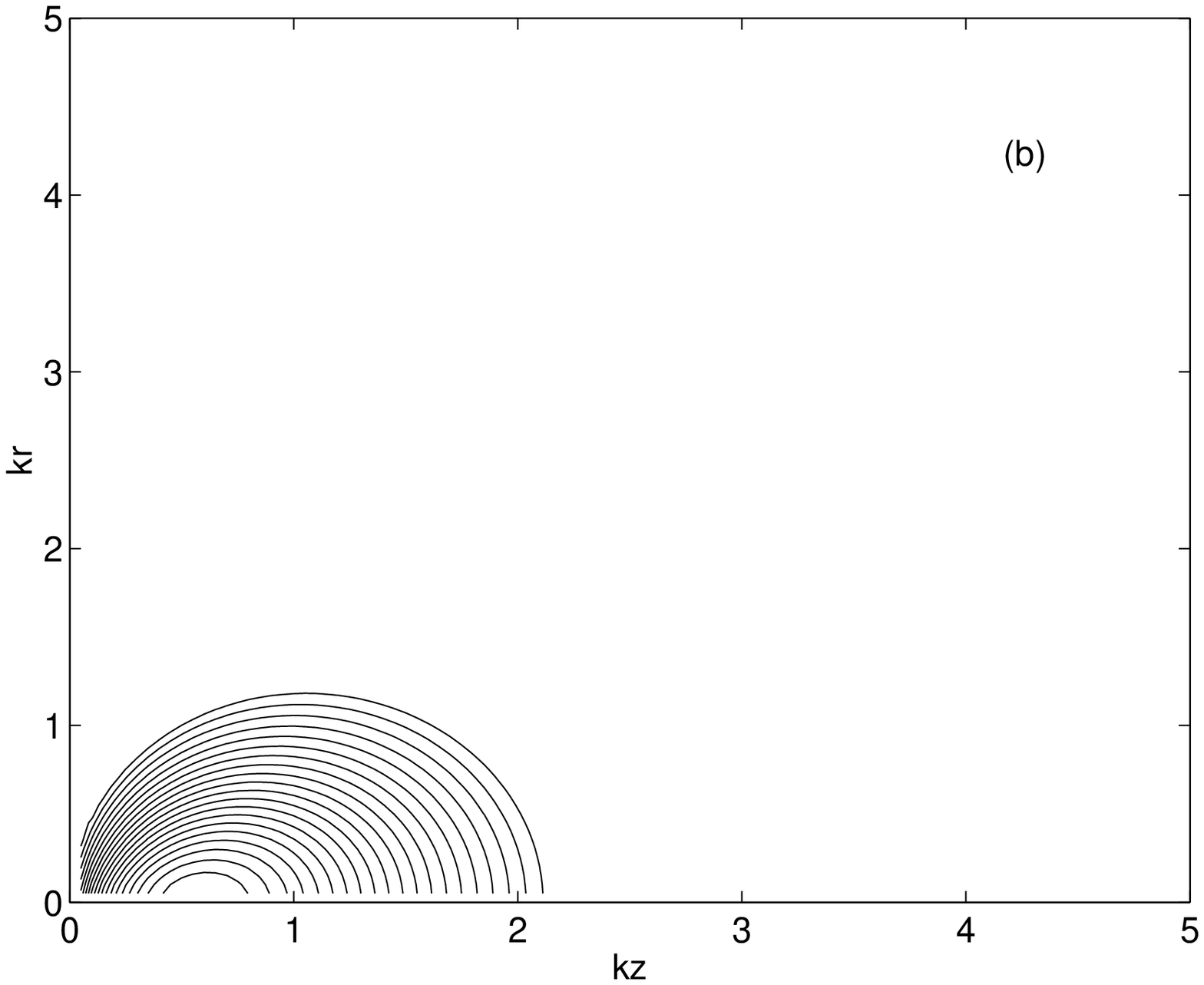}
\end{tabular}
\caption{Growth rates as a function of wavenumbers. An axial magnetic
field strength is fixed at $3000$ gauss for both experiments. 
The value of the maximum contour is $350 s^{-1}$ for (a) and $100 s^{-1}$
for (b). The
growth rate decreases with $k_r$ for both experiments. Up to $k_z=8$
modes are unstable in the sodium case (a), whereas only the $k_z=1$
mode is unstable in the gallium case (b). 
Other parameters are taken from Table~1.}
\label{fig4}
\end{figure*}
\begin{figure*}[hbt]
\begin{tabular}{cc}
\hspace{5mm}\includegraphics[width=8cm]{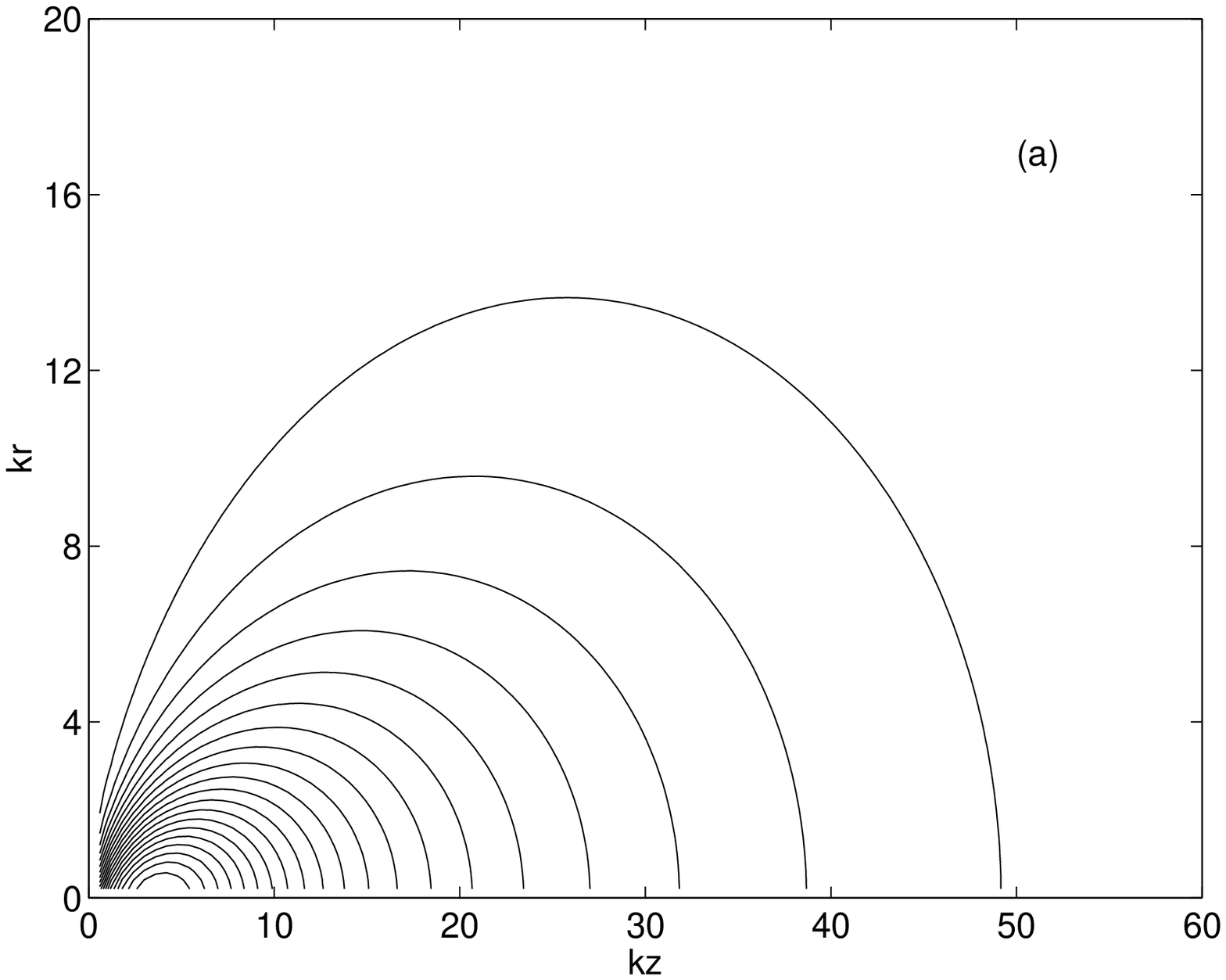}\hspace{5mm}
\includegraphics[width=8cm]{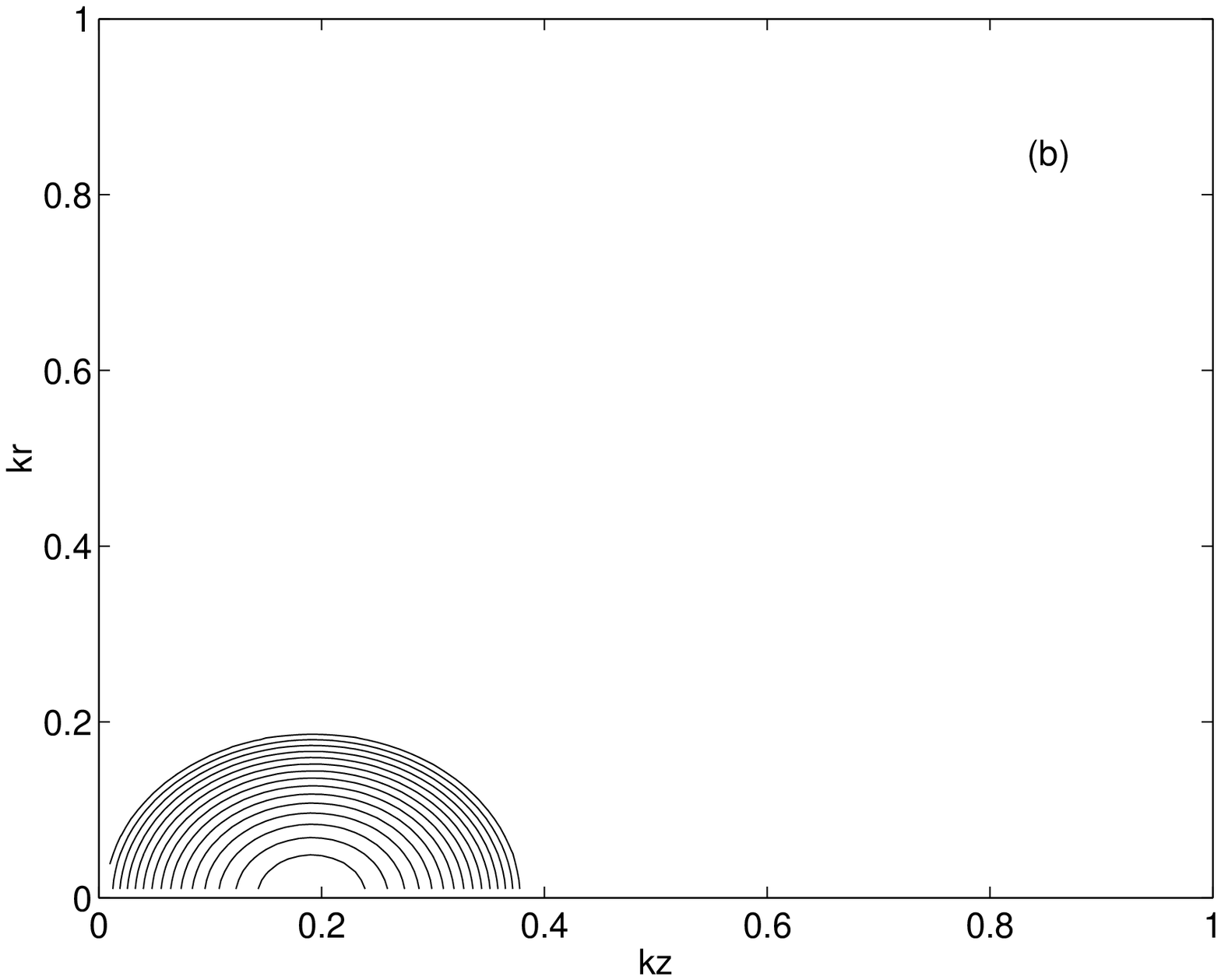}
\end{tabular}
\caption{Same as Fig.\ \ref{fig4} with $B_z=400$ gauss. The value of
  the maximum contour is $130 s^{-1}$ for (a) and $7.0 s^{-1}$
for (b). No mode is
destabilized in the gallium case (b), whereas high
$k_z$ modes are excited in the sodium case (a). }
\label{fig5}
\end{figure*}
\begin{figure*}[hbt]
\begin{tabular}{cc}
\hspace{5mm}\includegraphics[width=8.3cm]{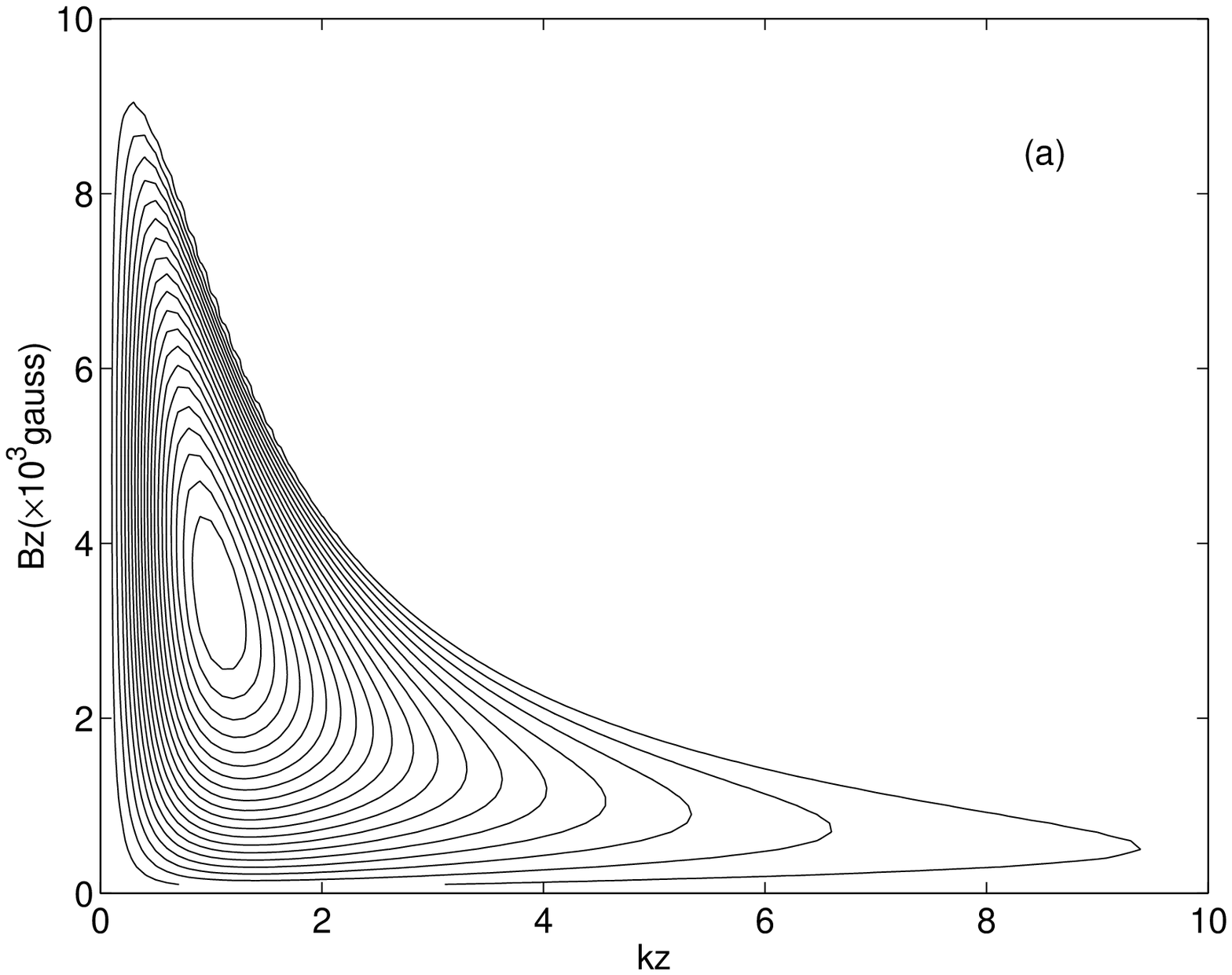}\hspace{5mm}
\includegraphics[width=8cm]{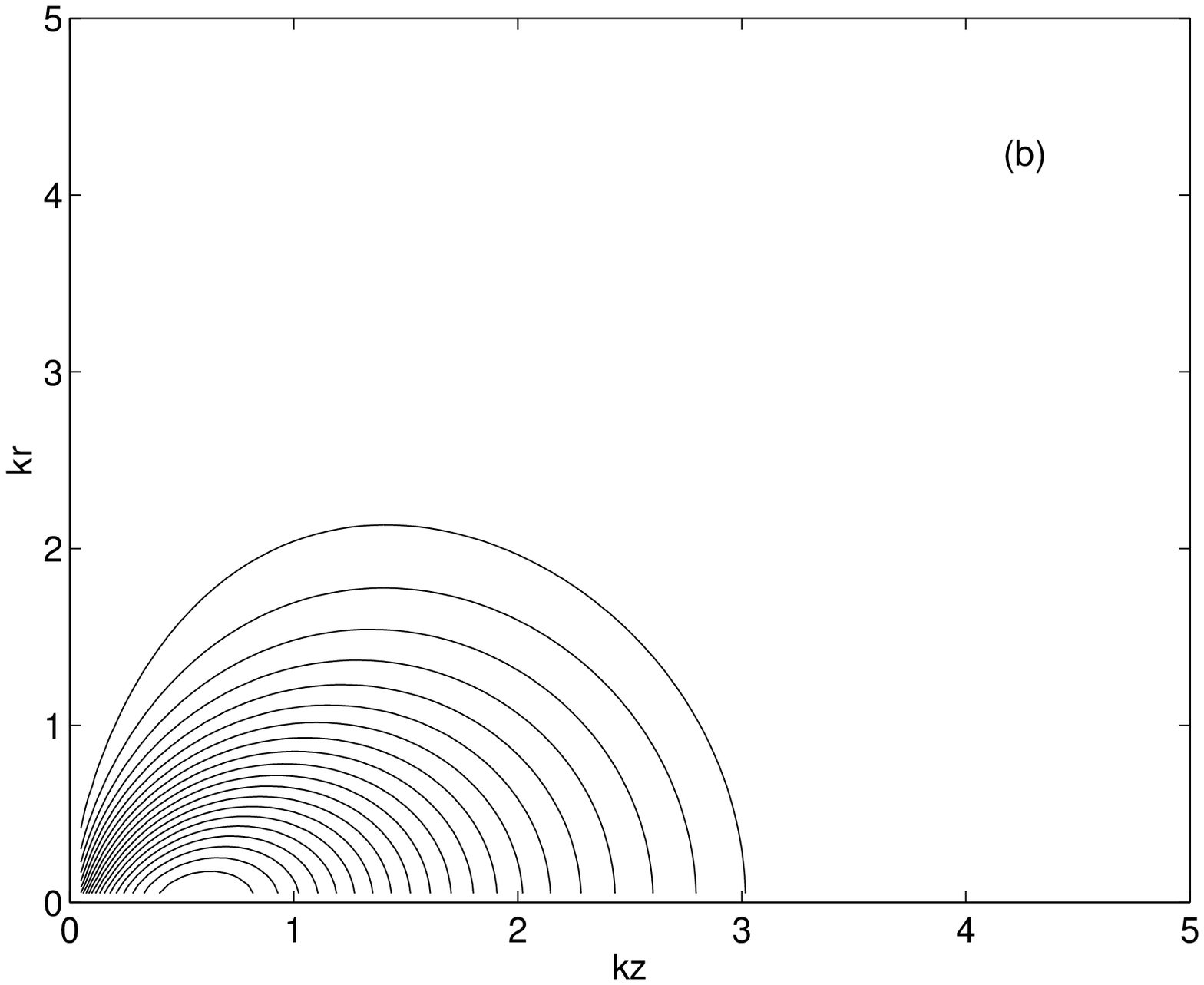}
\end{tabular}
\begin{center}
\includegraphics[width=8cm]{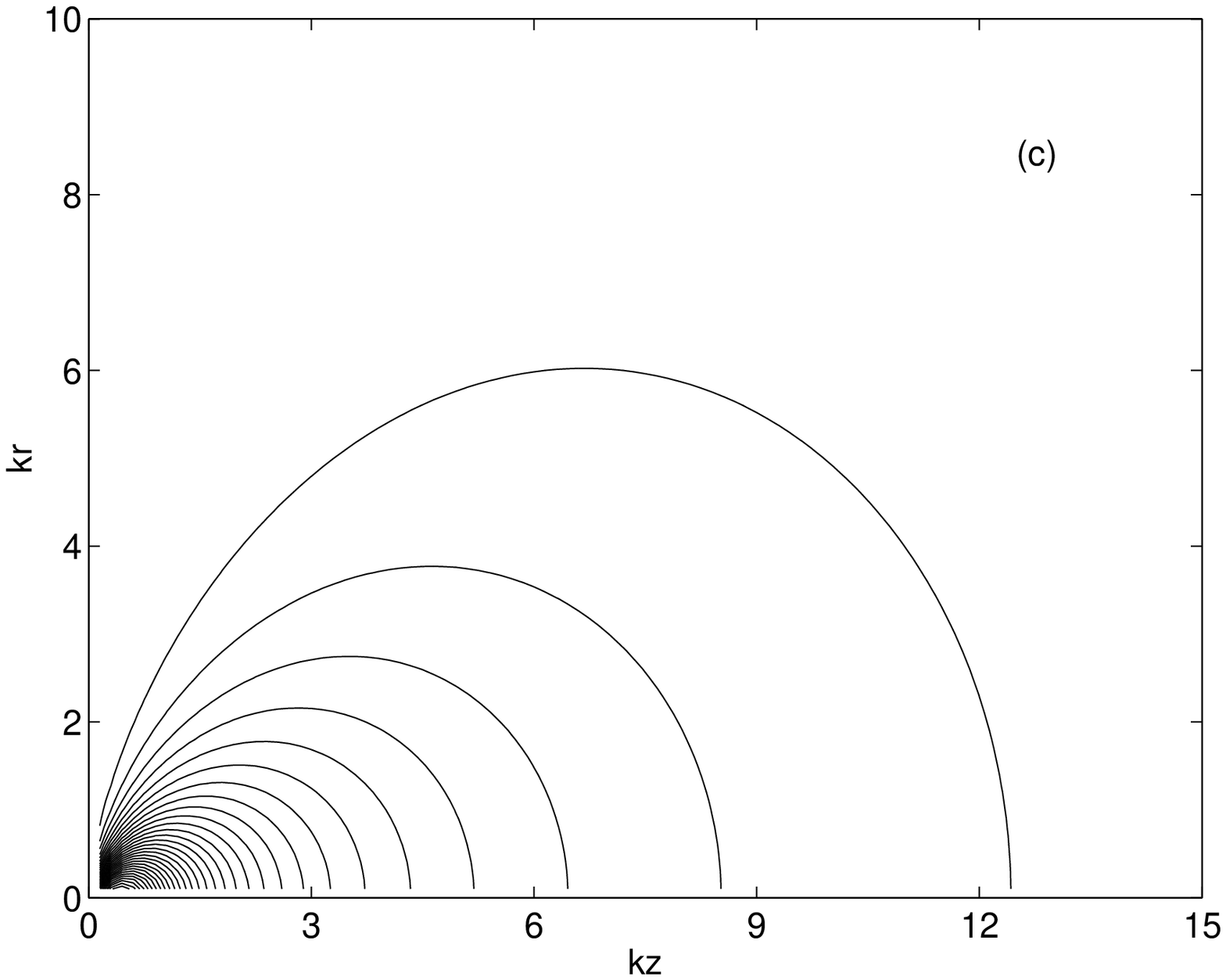}
\end{center}
\caption{Growth rates of the MRI in the gallium experiment with
$\kappa=0$. All the parameters are the same as the point C in
\citet{ji01} except
$\Omega_2=\Omega_1/9$. The value of the maximum contour is $35 s^{-1}$ for (a),
$100 s^{-1}$ for (b) and $30 s^{-1}$ for (c). (a): Growth
rate with $k_r=1$. When $B_z<4.5\times10^3$ gauss,
$k_z=2$ and higher modes are destabilized due to maximum shear
rate ($\kappa=0$). (b): With $B_z=3.0\times10^3$ gauss. High $k_r$
modes, which are stable in the finite
$\kappa$ case (Fig.\ \ref{fig4}b), are unstable. (c): With
$B_z=4.0\times 10^2$ gauss. High wavenumber modes ($k_r\leq15,
k_z\leq22$) are unstable, whereas no mode is unstable in the finite
$\kappa$ case (Fig.\
\ref{fig5}b).}
\label{fig51}
\end{figure*}

Using these parameters, the growth rate is obtained by solving
eq.~(\ref{dis}) numerically, and the unstable regions are plotted in
Fig.\
\ref{fig3} as a  function of axial magnetic field strength and wave
number $k_z$. The unit of $k_z$ is
$\pi/(R_2-R_1)$, $k_r$ is $\pi/L$ in Figs.\ \ref{fig3},
\ref{fig4},
\ref{fig5}, \ref{fig51}, \ref{fig10},
\ref{fig7} and \ref{fig6}.  The minimum possible values for the
dimensionless $k_r$ and $k_z$ are unity. We fixed $k_r$ as unity in
Fig.\ \ref{fig3}. Notice that in both experiments, a strong field
suppresses high $k$ modes because of the magnetic diffusivity and
magnetic tension.   In the sodium case (Fig.\ \ref{fig3}a), higher
$k_z$ modes are destabilized, and the growth rate is higher compared
to the gallium case (Fig.\ \ref{fig3}b).  In the gallium case, the
suppression of the unstable modes with low magnetic field occurs due
to finite
$\kappa$ (see Sec.\ 3 for details).

Figures~\ref{fig4} and \ref{fig5} demonstrate the dependence of the
growth rate on the wave numbers in the sodium and gallium
experiments.  In Fig.~\ref{fig4}, an axial magnetic field $B_z$ is
fixed at
$3\times10^3$ gauss, and in Fig.~\ref{fig5}, at
$400$ gauss. When $B_z=3\times10^3$ gauss, a number of
$k_z$ modes($k_z<8$) are destabilized in the sodium experiment(Fig.\
\ref{fig4}a), while in the gallium experiment(Fig.\ \ref{fig4}b) only
the $k_z=1$ mode is destabilized.

A more significant difference between the sodium and gallium
experiments is shown in Fig.\ \ref{fig5}. For weak magnetic fields no
mode is unstable in the gallium experiment(Fig.\ \ref{fig5}b), whereas
higher $k_z$ modes ($k_z>50$)  are excited in the sodium
experiment(Fig.\
\ref{fig5}a). As we noted before, the finite $\kappa$ suppresses the
unstable MRI modes with weak magnetic field.

In order to show the importance of the maximum shear, we plot the growth
rate for the gallium experiment with $\kappa=0$ in  Fig.~\ref{fig51}.
We reproduce  Figs.~\ref{fig3}b,
\ref{fig4}b and \ref{fig5}b in Figs.~\ref{fig51}a,
\ref{fig51}b  and \ref{fig51}c respectively, except we take
$\Omega_2=\Omega_1/9$ for the maximum shear [see Eqs. (\ref{ab}) and
(\ref{c})]. In Fig.\ \ref{fig51}a, high
$k_z$ modes are unstable with low
magnetic field, which are stable with finite
$\kappa$(Fig.\\ref{fig3}b). Maximum velocity shear also destabilizes high
$k_r$ modes (Figs.\ \ref{fig51}b and \ref{fig51}c),  so many modes will be
excited in the gallium experiment at maximum shear. In all cases, the
$(k_z, k_r)=(1,1)$ mode is dominant,  but mode coupling may occur in
the nonlinear regime to excite turbulence in the gallium  experiment.

\begin{figure*}[htb]
\begin{tabular}{cc}
\hspace{5mm}\includegraphics[width=8cm]{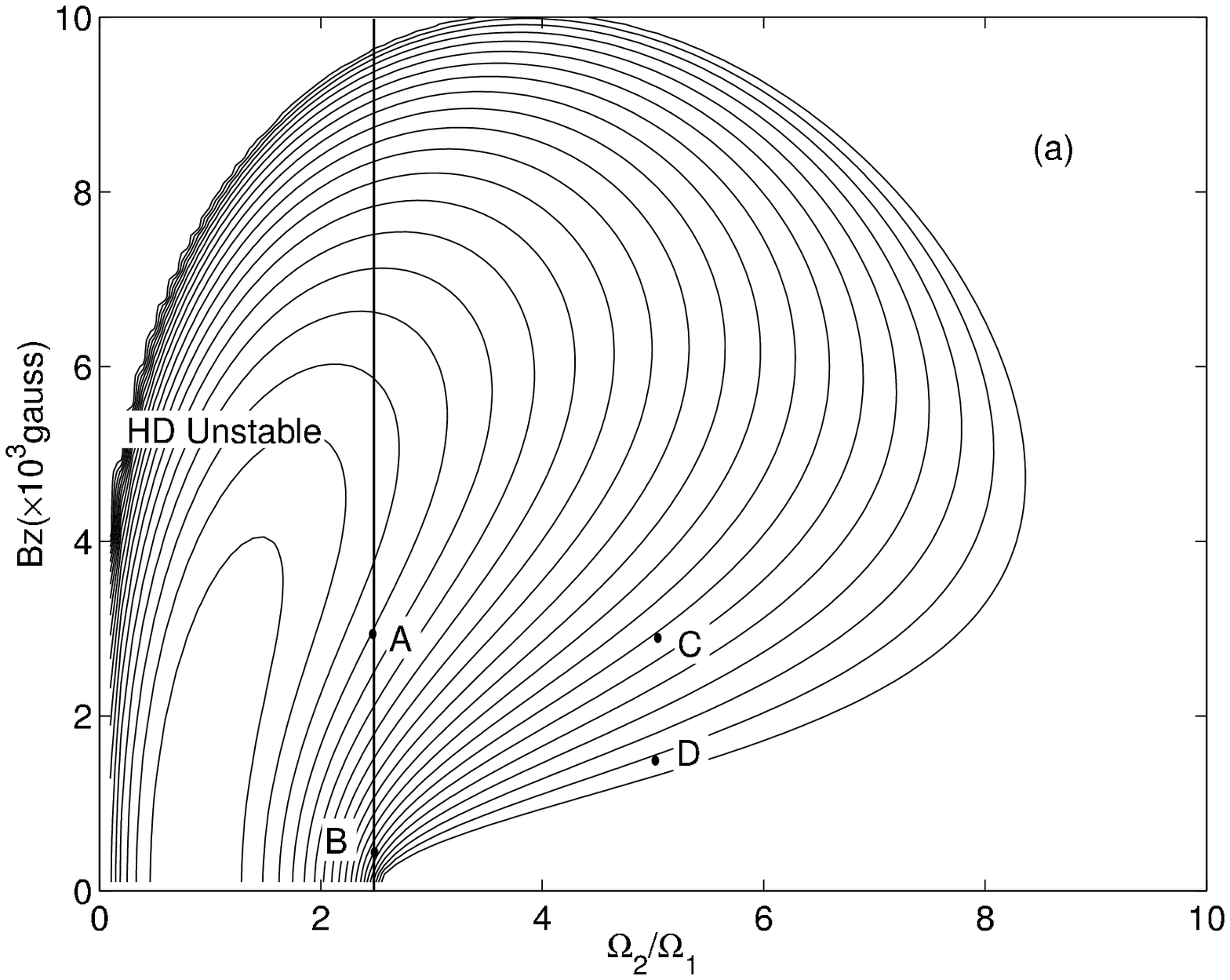}\hspace{5mm}
\includegraphics[width=8cm]{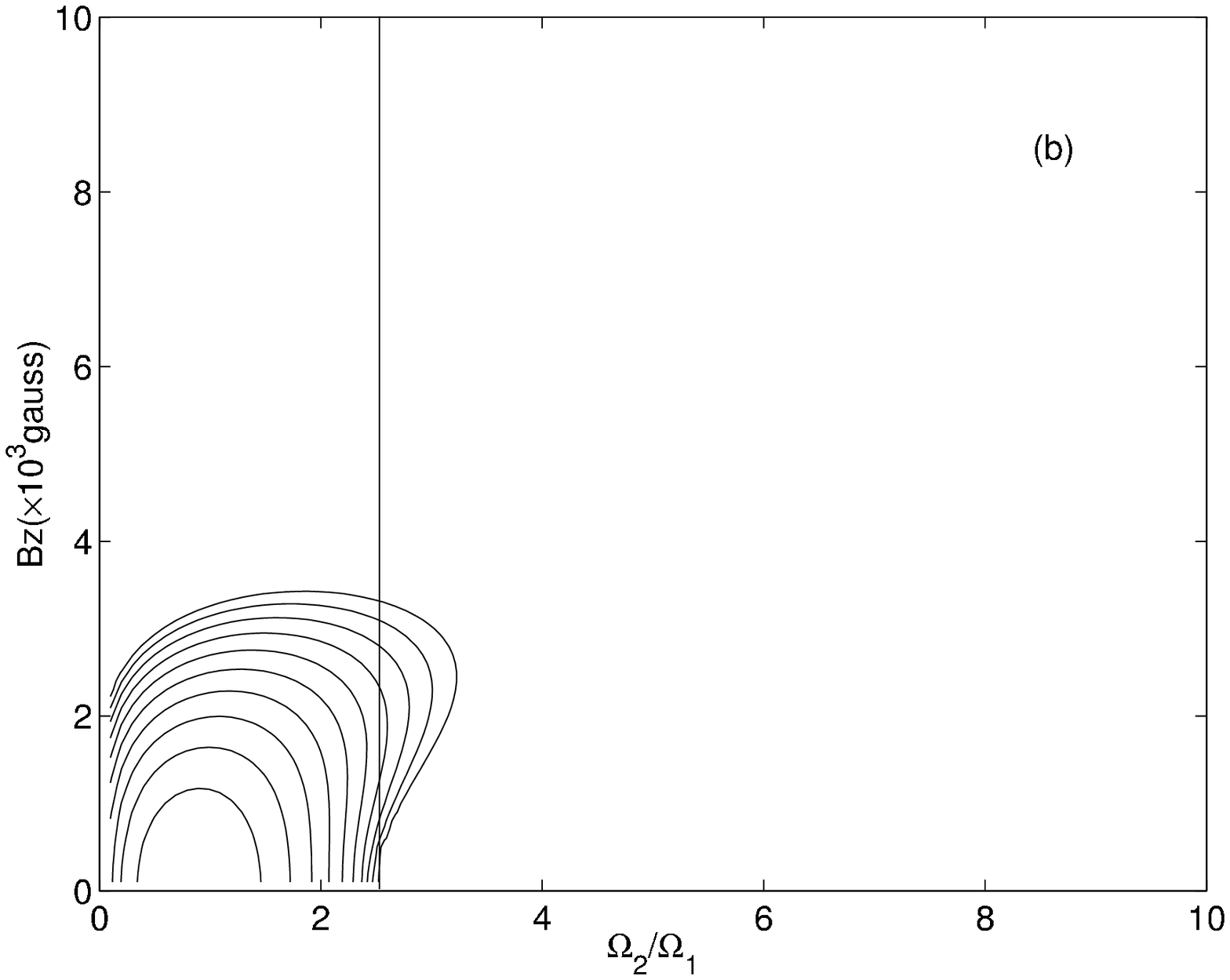}
\end{tabular}
\caption{The MRI unstable region of sodium experiment for two different modes.
Mode number is fixed as $(k_r ,k_z)=(1,2)$ (a)
and $(3,5)$ (b).  The value of the maximum contour is $270 s^{-1}$ for (a) 
and $250 s^{-1}$ for (b). The region
$\Omega_2/\Omega_1<0.25$ is hydrodynamically unstable. Parameters of
point A correspond  to those of Figs.\
\ref{fig3}a and \ref{fig4}a, B
   to those of Fig.\ \ref{fig5}a, C to Figs.\ \ref{fig10}a  and
\ref{fig10}b, D to Fig.\ \ref{fig10}c, respectively.  }
\label{fig9}
\end{figure*}
While the maximum shear flow
profile leads to easy excitation of MRI, it falls on the border line
for pure hydrodynamical instability.  We show the contour plot of the
unstable region for the modes
$(k_r, k_z)=(1, 2)$ and $(3,5)$ of the sodium experiment in Fig.\
\ref{fig9}. In the regime $\Omega_2/\Omega_1<0.25$, sodium is
hydrodynamically unstable. The maximum shear flow is indicated as a
solid line. High wavenumber modes are unstable only near the maximum
shear flow and weak magnetic field in the hydrodynamically stable region
(Fig.~\ref{fig9}b). 
\begin{figure*}[htb]
\begin{tabular}{cc}
\hspace{5mm}\includegraphics[width=8.1cm]{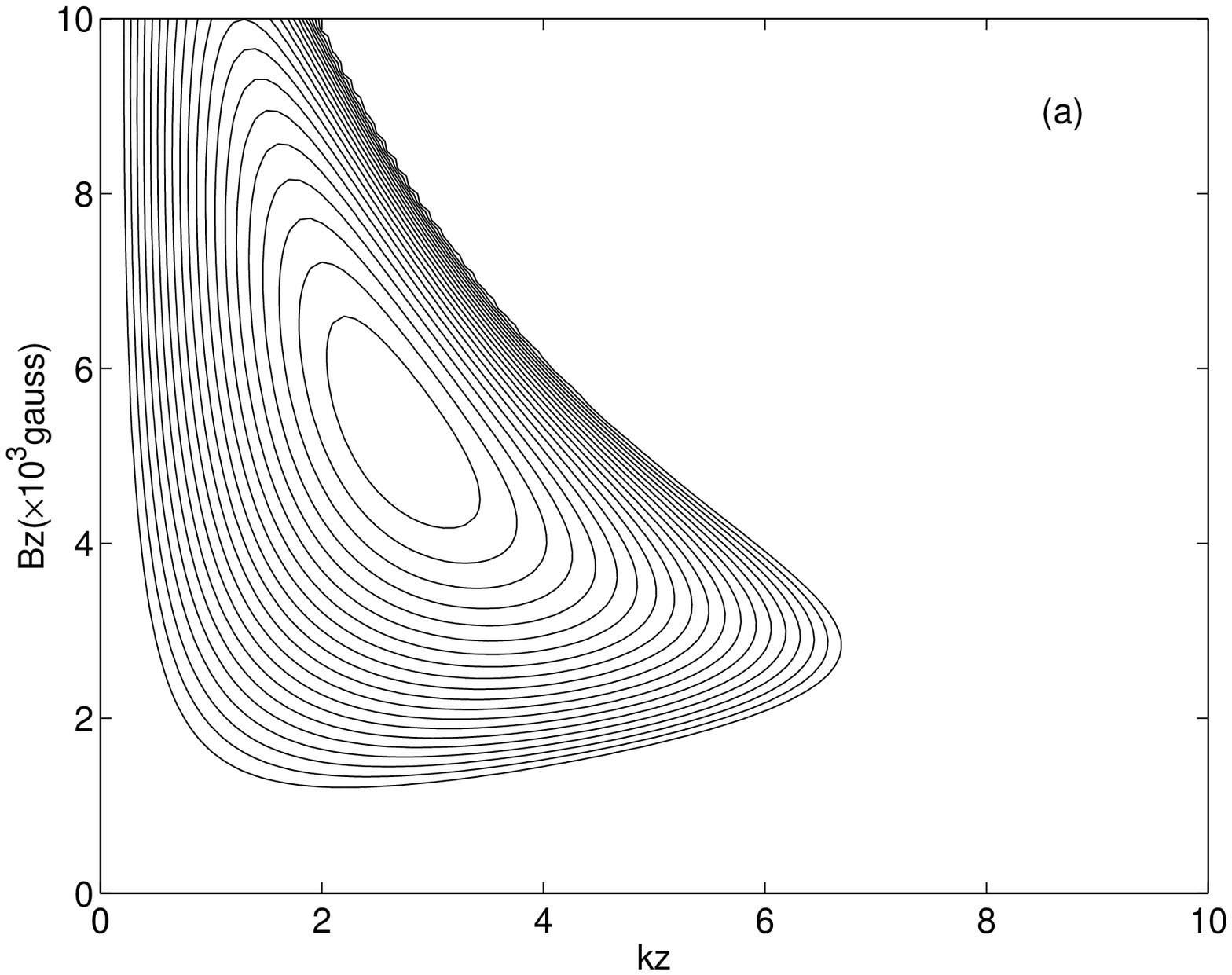}\hspace{5mm}
\includegraphics[width=8cm]{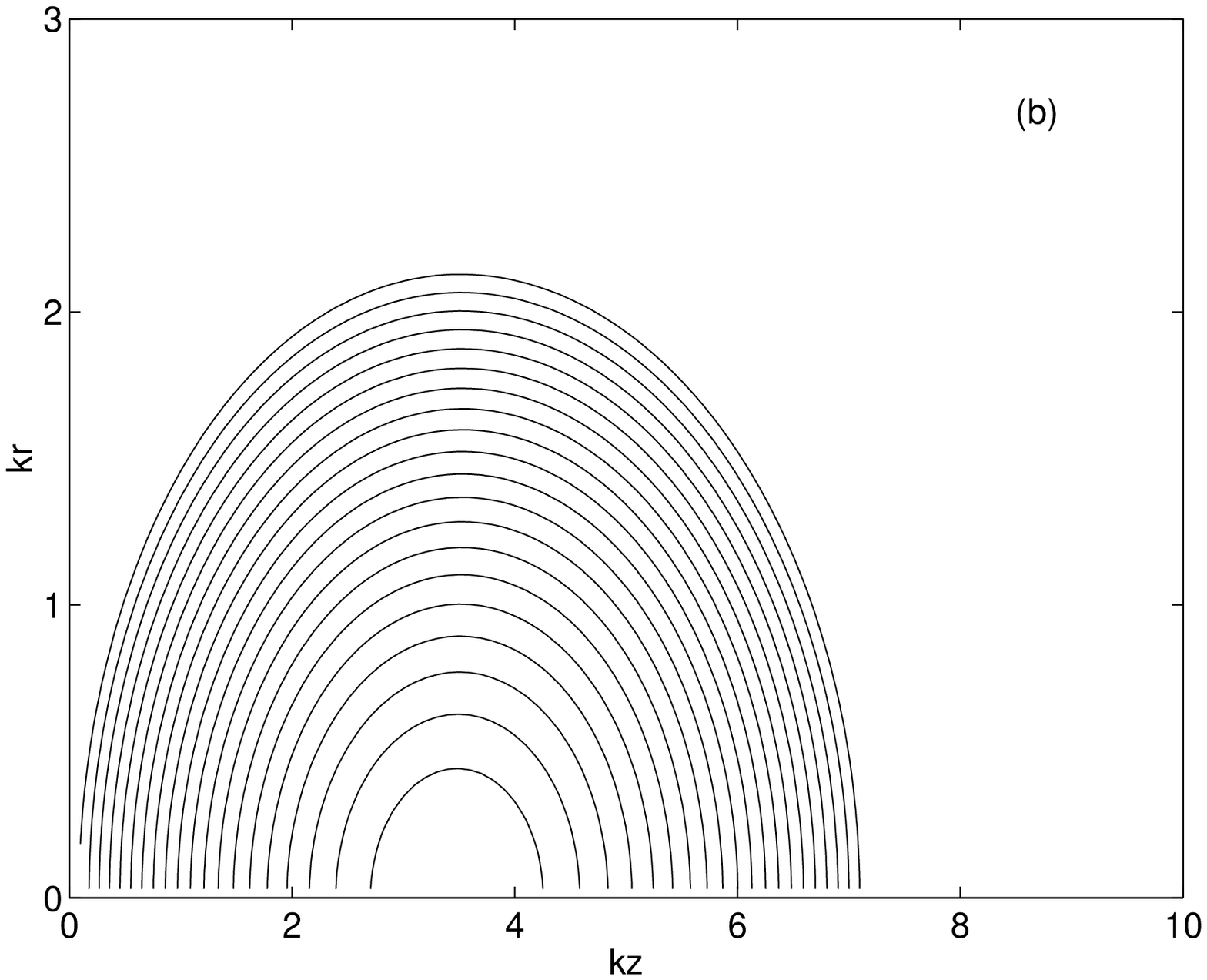}
\end{tabular}
\begin{center}
\includegraphics[width=8cm]{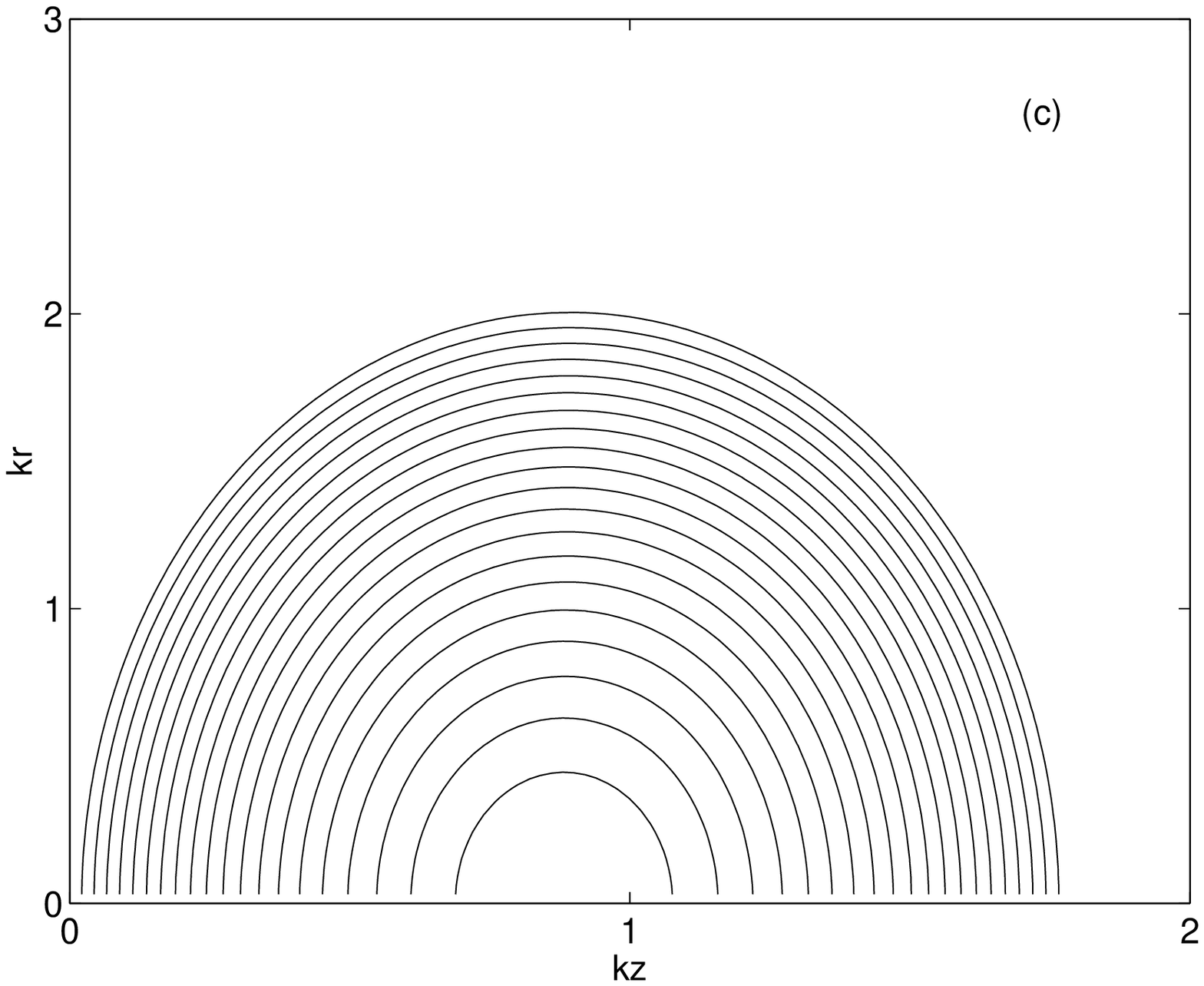}
\end{center}
\caption{The MRI growth rates with finite $\kappa$ flow shear,
   $\Omega_2/\Omega_1=0.5$,
for sodium experiment (Table~1). 
(a) The wavenumber $k_r$ is fixed as unity;
(b)With $B_z=3\cdot10^3$ gauss; (c)With
$B_z=1.5\times10^3$. The value of the maximum contour is $170 s^{-1}$ for (a)
and (b) and $65 s^{-1}$ for (c). 
High wavenumber modes are suppressed in each case. }
\label{fig10}
\end{figure*}
Growth rates for the finite $\kappa$ shear case are
shown in Fig~\ref{fig10}. High wave number modes are stabilized
compared to the
$\kappa=0$ case, Fig.~\ref{fig51}, but several modes are still
unstable.
Comparison of Fig.~\ref{fig3}b to
Fig.~\ref{fig10}a and Fig.~\ref{fig4}b to Fig.~\ref{fig10}b shows that
even for highly sub-critical flow with
$\Omega_1/\Omega_2=2$ the sodium experiment will allow one to observe more
MRI unstable modes than the gallium experiment for slightly
sub-critical flow with
$\Omega_1/\Omega_2=8.2$ (point C of Ji et al., 2001). Thus, the sodium
experiment has a higher potential for the observation of   turbulence
due to the nonlinear development of the MRI at the rotation profiles with
smaller shear, i.e. when the flow is highly stable in the absence of
the magnetic fields.

\section{Turbulence Derived from The Ekman Layer Flow }
\label{sec5}

We believe we need to understand the minimum expected  turbulence level in
the fluid before attempting to observationally separate the MRI growth
from an unknown background.  Of course that background will be measured
first, but this section deals with an estimate of its magnitude. First we
must point out that if there were no initial turbulence in the flow, then
the expectation of laminar flow is that the torque or power required to
drive the experiment would be negligibly small.  Instead, since we know
that there must be a torque associated with the Ekman  layer flow and
therefore a minimum power required to drive the Couette flow, we calculate
this and compare it to the observed instability
based upon the gradient of angular momentum
\citep{ric99}.

\citet{ric99} have extensively reviewed the earlier experimental
work on Couette flow of
\citet{wen33} \& \citet{tay36}.  In the maximally stable case where the
outer cylinder rotates at
$\Omega_2$  and the inner one is stationary, $\Omega_1 = 0$, and thus
where the angular momentum increases outward,  the flow is observed to be
weakly unstable  to a finite amplitude instability despite the prediction
of stability for a positive angular momentum gradient. This observed
instability has been invoked by
\citet{ric99} as a possible mechanism for the $\alpha$-viscosity of
Keplerian accretion disks.  The implication is that the free energy of the
flow  is accessed through  a finite amplitude instability producing
turbulence.  However, this turbulence in the positive angular momentum
gradient, theoretically  stable regime, is observed to be much weaker then
the inverse case,  exerting a much smaller torque than the turbulence
generated by the unstable Couette flow, which in turn is self-excited by
the flow of free energy through the turbulence itself.  This lack of
equivalence causes us to  ascribe the occurrence of
turbulence in the stable flow case to the back-reaction to the torque of the
Ekman layer flow.  We calculate this back reaction turbulence here as
the likely lower limit of turbulence for these experiments.
The initial experimental measurement of the torque as well as the in situ
magnetic fluctuations for very low fields can and will be compared to
these predictions.  Finally we note that the predicted ratio between the
torques  for the stable and unstable flows, based on the Ekman layer flow
$G_{\rm stable}/G_{\rm unstable} \simeq R_e^{1/2}/R_e$, 
roughly agrees with the measurements of \citet{tay36}. 
We delay until a later paper a full analysis
of this problem, but give here an estimate of this expected turbulent
viscosity as compared to the purely laminar one.

An Ekman layer forms adjacent to the surfaces of the end walls. This flow
is both radial and azimuthal, thin, laminar, and high speed. The resulting
flux of angular momentum creates a torque on the fluid and an unstable
velocity profile between the inner and outer cylinders. The unstable
shear flow at both the inner and outer cylindrical boundaries results in
a "law of the walls" or "logarithmic profile" turbulent boundary layer
\citep{sch60} with each of the cylinder walls. This turbulence extends
from the inner to the outer differentially rotating cylinders and
creates the turbulent stress necessary to transfer the torque between
them.

The assumption of the stability of Couette flow in the experiment is
limited by the formation of an  Ekman layer adjacent to the surfaces
of the end plates.  Since these end plates corotate with the outer
frequency $\Omega_2$, then at any radius $r \leq R_2$  the fluid will
be rotating faster than the end wall. An Ekman layer forms
\citep{prandtl52}, when the centrifugal force is not balanced by a
pressure gradient.
The pressure in the Ekman layer is
the same as the pressure in the bulk of the cylinder but
the centrifugal force is smaller in the Ekman layer because of friction with
the end wall.
As a result, a (negative) radial flow develops in
a thin layer of thickness $\delta$ with a mean radial velocity
$<v_{r}> \simeq  r \Omega/2$ while undergoing a mean azimuthal  motion
$<v_{\theta}> \simeq r \Omega/2$ .  The analysis of
\cite{prandtl52} results in the thickness $\delta
\simeq r/ \sqrt {R_{e}} = \sqrt{\nu/\Omega_2}\simeq 5.6
\times 10^{-3}$ cm in sodium and $7.0 \times 10^{-3}$ cm in gallium,
using the parameters of Table~1.  Hence a radial (negative) current,
$F_r$, flows of order
\begin{equation} F_r \simeq - \delta (2\pi R_2) R_2\Omega_2 /2 = -
\pi R_2^3 \Omega_2 /\sqrt {R_{e}} \, \, \, cm^3 s^{-1}.
\end{equation}

This (negative) radial flow at both ends towards the axis must be
balanced by a positive, slower radial flow throughout the central
region.  The Ekman flow merges with the central flow by a boundary
layer at the inner cylinder surface  resulting in a circulation within
the Couette flow volume driven by the Ekman layers at each end.  Since
this flow represents a flux of angular momentum, $\rho F_r R_2^2
\Omega_2 /2$, from the inner radius, $R_1$, to the outer radius,
$R_2$, there must be a  torque, $G$, transmitted by  the fluid
corresponding to the  difference in the  flux of angular momentum
between these two surfaces or
\begin{equation} G_E = \rho F_r  ( R_2^2 - R_1^2) \Omega_2 /2 = \rho (3
\pi/8) R_2^5
\Omega_2^2/\sqrt R_e.
\end{equation} where we have used the ratio $R_1/R_2 = 1/2$ for the
sodium experiment. (A factor of $5.9
\times 10^{-3}$ smaller is implied for the gallium experiment.) Not
only does this torque, between
$R_1$  and $R_2$ determine the power required  to drive the flow, but
also imposes a  requirement for a weak turbulence within the so-called
stable Couette flow in order to transmit this torque between $R_1$ and
$R_2$. This level of turbulence becomes the minimum effective
viscosity or  turbulent viscosity, $\nu_t$, of the MRI experiments.  By
way of comparison in addition  we calculate the torque as if the flow were
completely laminar and compare the two torques.

The shear stress for turbulent or laminar flow, $\tau_t$, $\tau_L$, is
characterized by either an effective turbulent viscosity,
$\nu_t$ or laminar, $\nu_L$.  In the turbulent case  the
angular momentum flux from the Ekman layer must be balanced by the viscous
stress from the rate of shearing,
$A = r d\Omega/dr = - 2\Omega_2 R_2^2 r^{-2}$ [see \citet{pri81}]
resulting in a viscous drag per unit area,
$\tau_t = \rho\nu_t  A  = - 2 \nu_t \Omega_2 R_2^2 r^{-2}
\rho$ and therefore a torque per unit length, $t_t = -2 \pi r^2
\tau_t  = -4 \pi  \nu_t \Omega_2 R_2^2\rho$. This has the proper
scaling since the torque must be independent of radius. The
corresponding laminar torque per unit length, $t_L = -2 \pi r^2
\tau_L  = -4 \pi  \nu_L \Omega_2 R_2^2\rho$ where the viscosity, $\nu_L$,
is fixed by the fluid properties and not variable with the strength of the
turbulence. Then the total laminar torque per  half length  becomes,
$G_L = t_L L/2 = 2 \pi \nu_L \Omega_2 L R_2^2\rho = 2 \pi L R_2^4
\Omega_2^2 / R_e$.  The ratio of the two torques  becomes
$G_L /G_E = (16/3) (L/R_2) R_e^{-1/2}$. Since $R_e$ is very large and
$L/R_2 = 1$,  the laminar torque is negligibly  small compared to the
Ekman  layer  torque and therefore torque balance requires turbulence to
enhance the effective viscosity.  This effective turbulent viscosity is
obtained   by equating the Ekman angular momentum flux to the viscous
shearing  torque giving
\begin{equation}
\nu_t = (3/16) \frac{R_2}{L} \frac{R_2 \Omega_2}{\sqrt R_e}  R_2 =
10.0 \,\,\& \,\, 12.7 \,\, cm^2 s^{-1},
\label{eqn_nu_t}
\end{equation}  
for the sodium and gallium experiments respectively.

The structure  of this turbulence is problematic. It has been described by
\citep{tay36}, as initially a series of long parallel vortices,
"Taylor columns"  that extend the full length of the annular space and
that at greater $R_e$  these columns break up becoming fully developed
turbulence.  We expect at some value of magnetic field strength that these
vortices  will be suppressed by magnetic
field of sufficient strength.  It seems unlikely,  however,  that a return
to laminar flow would take place, because the flow profile, with no
turbulent shear stress, but still the Ekman flow, will become more
distorted from the stable profile resulting in stronger turbulent drive.
However,  for now we defer analysis and expect guidance from  future
experiment.

Finally a   turbulent magnetic Prandtl number,
$P_{Mt}$, for measuring the strength of this Ekman driven  turbulence,
can also be estimated as
\begin{equation} P_{Mt} = \frac{\nu_t}{\eta}= 0.012 \,\, \&
\,\, 6.3 \times 10^{-3}
\end{equation} 
respectively. Despite the very small size of the  Ekman
layer, the turbulence generated by such a flow influences the ability to
distinguish turbulence  caused by the MRI at low values of magnetic
field from the  hydrodynamic turbulence caused by the Ekman layer. At
higher values of the magnetic field, above that affected by the Ekman
turbulence,   the effects caused by the MRI should be clearly
recognizable.

\begin{figure*}[htb]
\begin{tabular}{cc}
\hspace{5mm}\includegraphics[width=8cm]{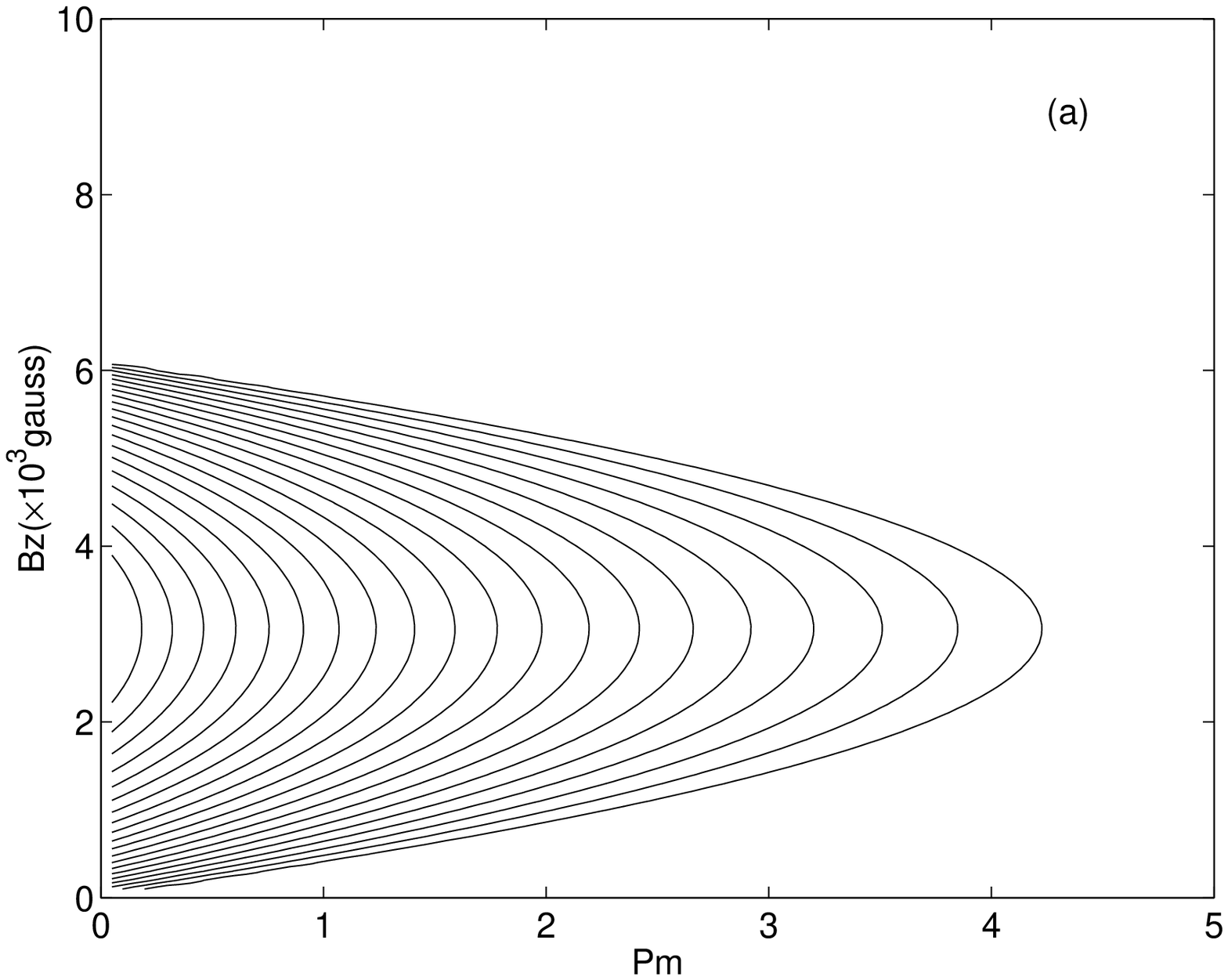}\hspace{5mm}
\includegraphics[width=8cm]{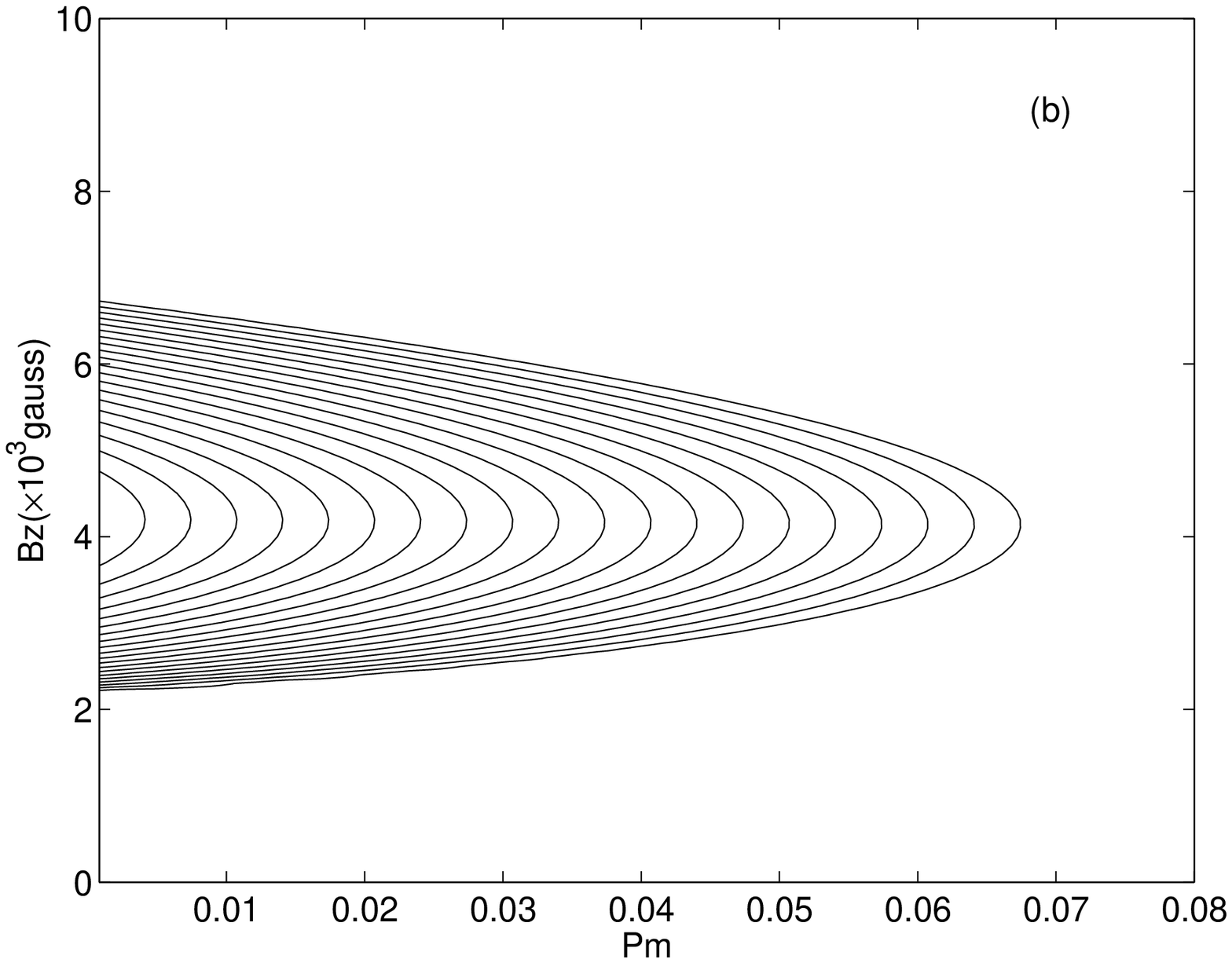}
\end{tabular}
\caption{The effect of turbulence on the MRI for the sodium (a) and
gallium (b) cases. The growth rate is plotted as a function of an axial
magnetic field and a magnetic Prandtl number. Wavenumbers are fixed
for the maximum growth rate 
of each experiment [$(k_r,k_z)=(1,4)$ for sodium, $(1,1)$ for
gallium]. 
The value of the maximum contour is $280 s^{-1}$ for (a) and $23 s^{-1}$ for
(b). The growth rate
decreases when the turbulent viscosity increases. The MRI can be
sustained in fully turbulent fluid in both cases for the minimum Ekman
layer driven turbulence, ($P_m=0.012$ for sodium,
$P_m=6.3\times10^{-3}$ for gallium).}
\label{fig7}
\end{figure*}
\begin{figure*}[htb]
\begin{center}
\includegraphics[width=8cm]{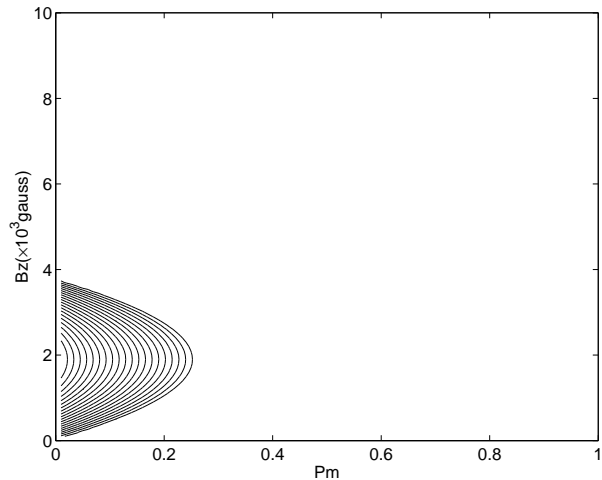}
\end{center}
\caption{The effect of the turbulence on the MRI for the mode
$(k_r,k_z)=(3,4)$ in the sodium experiment. The value of the maximum contour is
$90 s^{-1}$. Note that this mode is
stable in gallium even without turbulence. This mode will remain
unstable even with turbulence. The critical Prandtl number for this
mode to become stable is
$P_m\simeq0.3$. }
\label{fig6}
\end{figure*}
Figure~\ref{fig7} shows the dependence of the
MRI growth rate with the turbulent
$\nu_t$ as a function of an axial magnetic field strength and the
Prandtl number for the most unstable modes in both experiments. In
sodium experiment, higher $k_r$ modes are also unstable. Figure
\ref{fig6} shows the growth rate for
$(k_r,k_z)=(3,4)$ mode. Note that this mode is stable in gallium, even
with laminar viscosity only. Finite $\kappa$ for the gallium
experiment prevents the MRI from developing with weak magnetic
field(compare Figs.~\ref{fig3}b and
\ref{fig51}a, Figs.~\ref{fig4}b and~\ref{fig51}b above),  and MRI
exists only in the region
$3\times10^3\,\mbox{gauss}<B_z<6\times10^3\,\mbox{gauss}$
(Fig.~\ref{fig3}b). However, the sodium experiment with maximum shear,
$\kappa=0$, can be destabilized with already very weak magnetic field,
in the range of
$50\,\mbox{gauss}<B_z<100\,\mbox{gauss}$.

In conclusion, the presence of Ekman layers is significant
for the  determining of the power necessary to sustain the
differential rotation in the apparatus but has a negligible effect on
the condition of the  excitation of the MRI as it has been already
mentioned briefly in  Ji et al.~(2001). However, the turbulence
excited due to the presence  of Ekman layers may interfere with our
measurements of perturbations of magnetic field excited by the MRI.
Therefore, the
presence of weak turbulent perturbations due to  Ekman layers seems
unavoidable whenever one observes the excitation  of the MRI. It also seems
unlikely that the imposed magnetic field in both sodium and gallium
experiments can significantly exceed a value of a  few thousands
Gauss. The characteristic amplitude of the perturbations of the magnetic
field due to the Ekman layer turbulence is
$\sim B_z\lambda v_t/\eta \approx P_{Mt} B_z$. The typical value of
such perturbed magnetic fields is of order of 1~\% of the applied
field  (see Table~1 for $P_{Mt}$), thus, limiting the possible  MRI
measurements of growing fields to more than 1~\% of  the initial
magnetic field.

\section{Discussion and Conclusions}\label{sec6}

There are several aspects of each experiment that warrant discussion.
The  first deals with the analysis performed for both experiments.  In
this  paper we used the full dispersion relation which depended not
only on an azimuthal magnetic field but also has all terms
proportional to
$1/r$ retained.  This corresponds  to the geometrical effect of the
curvature from  the cylindrical geometry.  Only when one neglects all
$1/r$ terms in the  dispersion relation does one obtain  the results
of \citet{ji01}. This allows us to consider many different magnetic
field configurations, some of which will be suitable for studying the
MRI.

Nevertheless, even with these significant problems and differences, it
should be noted that both the NMD and PPPL experiments have an
excellent chance of  observing the MRI in the laboratory.  Both
experiments obtain very high  growth rates under varying conditions
yielding a flexible set of opportunities.

Finally we note that the effect of an azimuthal magnetic field and
an analysis of nonaxisymmetric modes are still open problems.
\citet{nog00} showed that the local dispersion analysis in shear flow
may fail even for the qualitative estimation of growth rates. The
eigenmode analysis for nonaxisymmetric modes is necessary for further
understanding of the  MRI instability in the NMD experiment.
We are now developing the shooting method code
for solving Eqs.~(\ref{b1})-(\ref{b4}) simultaneously. Spatial
dependence of the radial wavenumber and azimuthal magnetic field
dependence will be analyzed.

\acknowledgments K.N. and S.A.C. are particularly indebted to Hui Li
of Los Alamos National Laboratory for pointing out the relevance of the
MRI to the NMD experiment and encouraging the present work.
V.P. thanks Eric Blackman for stimulating
conversations and acknowledges partial support from DOE grant
DE-FG02-00ER54600. 

We all acknowledge important comments by the referee, which significantly
improved the article.
In addition this
work has been supported by  the DOE, under contract W-7405-ENG-36.

\end{document}